\documentclass[
  reprint,
  amsmath,amssymb,
  pra,
  aps
]{revtex4-2}
\usepackage{graphicx}
\usepackage{hyperref}
\usepackage[linesnumbered, ruled, vlined, titlenotnumbered, noend]{algorithm2e}
\usepackage{float}
\usepackage{amsmath, amsfonts, amssymb, amsthm, color}
\usepackage{tikz-cd}
\usepackage[capitalise]{cleveref}
\crefname{algocf}{alg.}{algs.}
\Crefname{algocf}{Algorithm}{Algorithms}
\usepackage{subcaption}
\usepackage{braket}

\usepackage{tabularx}
\usepackage{makecell}
\usepackage{multirow}

\usepackage[T1]{fontenc}

\usepackage{appendix}

\newcommand{\plog}{\varepsilon_L}

\newcommand{\FTwo}{{\mathbb F}_2}

\newcommand{\mA}{{\cal A}}
\newcommand{\mB}{{\cal B}}

\newcommand{\BaIon}{^{133}\text{Ba}^{+}}
\newcommand{\Sqb}{6 ^{2}S_{1/2}}
\newcommand{\Dqb}{5 ^{2}D_{5/2}}
\newcommand{\Pqb}{6 ^{2}P_{1/2}}
\newcommand{\Ttwo}{T_2^{*}}
\newcommand{\Tone}{T_1}

\newcommand{\Txsec}{T_X}
\newcommand{\FidAvg}{F}

\newcommand{\Hxred}{H_X^{\text{red}}}
\newcommand{\Hzred}{H_Z^{\text{red}}}
\newcommand{\hxred}{h_X}
\newcommand{\hzred}{h_Z}
\newcommand{\Noise}{{\cal N}}
\newcommand{\Noiset}{\Noise(t)}

\newcommand{\nSEC}{n_\text{SEC}}

\setcitestyle{super} 
\begin{document}

\title{Breakeven demonstration of quantum low-density parity-check codes}
\author{Edwin Tham}
\thanks{These authors contributed equally to this work.}
\affiliation{IonQ, Inc.}

\author{Michael L. Goldman}
\thanks{These authors contributed equally to this work.}
\affiliation{IonQ, Inc.}

\author{Shantanu Debnath}
\affiliation{IonQ, Inc.}

\author{Ashay N. Patel}
\affiliation{IonQ, Inc.}

\author{Moiz Rasheed}
\affiliation{IonQ, Inc.}

\author{Erik Nielsen}
\affiliation{IonQ, Inc.}

\author{Matthew Boguslawski}
\affiliation{IonQ, Inc.}

\author{Jonathan Mizrahi}
\affiliation{IonQ, Inc.}

\author{Jason Nguyen}
\affiliation{IonQ, Inc.}

\author{Jyothi Saraladevi}
\affiliation{IonQ, Inc.}

\author{Felix Tripier}
\affiliation{IonQ, Inc.}

\author{Adam West}
\affiliation{IonQ, Inc.}

\author{Min Ye}
\affiliation{IonQ, Inc.}

\author{Neal Pisenti}
\affiliation{IonQ, Inc.}

\author{Kenneth Wright}
\affiliation{IonQ, Inc.}

\author{John Gamble}
\affiliation{IonQ, Inc.}

\author{Nicolas Delfosse}
\affiliation{IonQ, Inc.}

\date{\today}

\begin{abstract}
High-rate quantum low-density parity-check (qLDPC) codes are a leading candidate for fault-tolerant quantum computing. They feature higher encoding rates than planar alternatives such as the surface code, but their implementation often entails significant hardware hurdles like the need for long-range couplers.
We leverage the flexibility of a trapped-ion quantum computer to demonstrate eight quantum error-correcting codes with starkly different qubit connectivity requirements on a single device without any hardware reconfiguration.
These experiments span three families of quantum error-correcting codes: qLDPC codes, topological codes, and concatenated codes.
With a qLDPC code encoding 4 logical qubits into 18 physical qubits, we achieve a logical error rate up to $9\times$ better than a previous demonstration of a similar code on superconducting solid-state qubits.
Moreover, our implementation exhibits breakeven performance, with some instances achieving qubit lifetimes comparable to or slightly exceeding   that of our trapped-ion qubits.
We use a novel implementation of the optical-metastable-ground (OMG) architecture for addressable mid-circuit measurement and reset, which enables us to perform these experiments without any ion transport or dedicated coolant ions, requirements that typically consume a large fraction of the runtime or ion count of trapped-ion quantum computers.
\end{abstract}

\maketitle

\section{Introduction}
\label{sec:Intro}
To reach large-scale quantum computing applications, 
a fault-tolerant quantum computer (FTQC) is needed that integrates extensive quantum error correction~\cite{shor1996fault}. The surface code is one of the most popular choices because it can be implemented with a two-dimensional grid of qubits using only gates acting on neighbouring qubits~\cite{dennis_topological_2002, fowler_surface_2012}.
Leveraging the simplicity of this planar layout, the surface code was realized experimentally by several teams~\cite{krinner2022realizing, zhao2022realization, bluvstein_logical_2024, acharya_quantum_2025, he2025experimental}, culminating in Google's experiment below the breakeven point, which produced a logical qubit whose lifetime is longer than the lifetime of its constituent physical qubits~\cite{acharya_quantum_2025}.
However, the qubit overhead of the surface code is massive, as each logical qubit is expected to consume a block of hundreds of physical qubits to reach a sufficiently low logical error rate for large-scale applications.

Quantum low-density parity-check (qLDPC) codes~\cite{mackay2004sparse, breuckmann2021quantum} are obtained by relaxing the locality constraints of the surface code, allowing qubits to interact beyond nearest neighbours as long as each qubit is interacting with a bounded number of other qubits. This results in codes that encode a larger number of logical qubits per block, reducing the qubit overhead of quantum error correction.
Numerical simulations show that several families of qLDPC codes can outperform surface codes, and FTQC architectures relying on qLDPC codes are discussed in~\cite{xu2024constant, yoder2025tour, webster2026pinnacle, cain2026shor, tripier2026fault}. Promising qLDPC codes include hypergraph product (HPG) codes~\cite{tremblay2022constant, aydin_cyclic_2026}, bivariate bicycle (BB) codes~\cite{bravyi_high-threshold_2024} and their BB5 variant~\cite{ye_quantum_2025}, and radial codes~\cite{scruby2026high}.
However, experimental realization of these codes is non-trivial due to the challenge in implementing non-local gates.
Only recently has one particular instance of a BB code been shown experimentally, which used a superconducting chip equipped with long-range couplers built to implement the specific BB code~\cite{wang_demonstration_2026}.
This experiment realized a BB code encoding $4$ logical qubits in $18$ physical ones, using $32$ superconducting transmon qubits, achieving a logical error rate of $\approx 9\%$ per logical qubit per syndrome extraction cycle.

In this work, we use a trapped-ion device to demonstrate five distinct high-rate qLDPC codes.
On a BB5 code with the same number of logical and physical qubits as one used in a previous BB code demonstration~\cite{wang_demonstration_2026}, we achieve a logical error rate $4 \times$ better for phase-flip errors and $9\times$ better for bit-flip errors. 
We also report logical performance in the breakeven regime.
Defining a qubit lifetime as the $1/e$ survival time averaged over all single-qubit state preparations, several code instances exhibit logical qubit lifetimes comparable to, and in one case marginally exceeding, that of the underlying physical qubits.
In one code, the lifetime exhibited was $3.95\pm 0.68$~s, compared to $3.84\pm0.48$~s for our physical qubits.

To underscore the flexibility of our hardware, we also implemented two other classes of codes with larger encoding rates than the surface code: topological codes on a manifold with toroidal topology and concatenated codes.
We experimentally demonstrate toric codes, encoding two logical qubits per block, and the $[[4,2,2]]$ code concatenated with itself, producing a code encoding 4 logical qubits into 16 physical qubits. 
Trapped-ion devices had previously been used to demonstrate various block codes~\cite{ryan-anderson_implementing_2022, egan_fault_tolerant_2021, ryan2024high, paetznick_demonstration_2024, reichardt2024demonstration, aasen2026}, and larger concatenated codes are demonstrated in~\cite{rines2025demonstration, dasu2026computing}.

The performance and flexibility of our device stem from several key attributes.
First, all our gates are based on steerable Raman beams addressing any ion -- or pairs thereof -- in a single stationary chain \cite{Chen2024benchmarkingtrapped}, obviating the need for physical transport.
Second, we use a novel implementation of the optical-metastable-ground (OMG) architecture~\cite{allcock_omg_2021}, which allows for in-place mid-circuit measurements (MCM) of many qubits in each MCM round.
Notably, our OMG implementation allows us to use ancilla qubits, which are measured frequently via MCM in any QEC implementation, to also sympathetically cool the chain.
This avoids a need for dedicated coolant ions and sidesteps significant technical complexities in co-trapping different atomic species.
Finally, by not using physical transport at all, we have far more modest cooling requirements.
By contrast, in trapped-ion systems that rely heavily on transport of ions, transport and cooling consume a majority of execution time\cite{ransford2025helios}, and up to 50\% of the ions are used only for cooling~\cite{moses2023race, ransford2025helios}.

\section{Mid-Circuit Measurement}
\label{sec:methods}

\begin{figure*}[t]
\begin{centering}
\includegraphics[width=\linewidth]{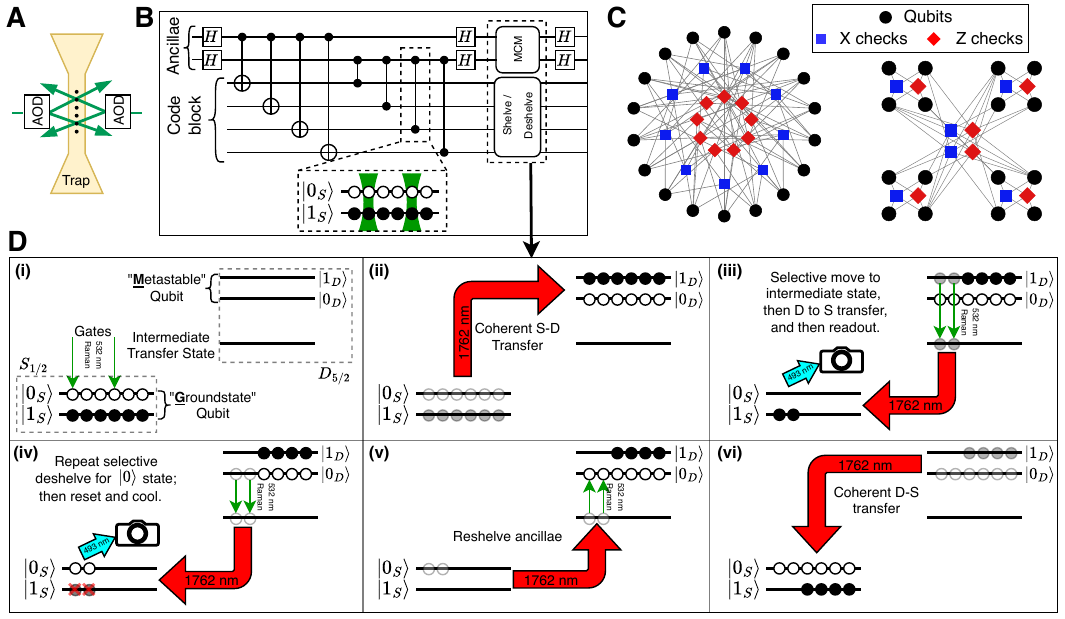}
\par\end{centering}
\caption{\textbf{A} Ion trap depicted with six ions, with gates realized via counter-propagating Raman beams (green) steered from either side with acousto-optic deflectors (AOD). \textbf{B} An example syndrome circuit, measuring $XXXX$ and $ZZZZ$ stabilizers of the $[[4,2,2]]$ code using two ancillae (top rails) prepared in $\ket{+}$ and measured in $X$.
\textbf{C} Tanner graphs whose edges depict the highly non-trivial two-qubit gate connectivities required to implement syndrome circuits for a BB5 (left) and concatenated $[[4,2,2]]$ (right) codes.
\textbf{D} Sequence of physical operations implementing OMG-based mid-circuit measurements, with open (filled) circles representing the 0 (1) qubit state.}
\label{fig:MainFig}
\end{figure*}

Our implementation uses a trapped-ion quantum computer based on a stationary chain of forty $\BaIon$ ions (\cref{fig:MainFig}A).
Each physical qubit is encoded in the states $\ket{0_S}\equiv\ket{F=0,m_F=0}$ and $\ket{1_S}\equiv\ket{F=1,m_F=0}$ of the ground $\Sqb$ manifold.
\Cref{fig:MainFig}D depicts the standard sequence of physical operations in our system.
In step (i), we apply sequential, unitary one- and two-qubit gates by illuminating target ions in $\Sqb$ with 532~nm beams (green).
This architecture features all-to-all connectivity without qubit transport.

The sequence from (ii)-(vi) shows our implementation of MCM using the OMG architecture. This architecture utilizes phase-coherent shuttling of a qubit between multiple encoding manifolds within a single ion, enabling MCM without the need to shuttle ions spatially. It also enables sympathetic cooling of the chain's motional modes without dedicated coolant ions or a separate atomic species, which constitute up to half of the ions in other trapped-ion devices~\cite{ransford2025helios}.

To protect idle qubits not measured during MCM, we first {\em shelve} the whole chain by using a global, bichromatic 1762~nm beam (red) to transfer all ions from $\ket{0_S}$ and $\ket{1_S}$ to the states $\ket{0_D}\equiv\ket{F=2,m_F=+1}$ and $\ket{1_D}\equiv\ket{F=3,m_F=+1}$ in the metastable $\Dqb$ manifold.
These states were chosen because their energy difference is first-order insensitive to magnetic fluctuations around our system's bias field of $17.7$~G.
We then selectively deshelve some subset of ions to be measured, termed the {\em readout ancillae}, through combined use of the same steerable Raman and global 1762~nm beams.
Raman beams sequentially drive transitions on specific ions between one of the qubit states and an intermediate state in $\Dqb$, and then the 1762~nm beam deshelves all population in that intermediate state in parallel across all ions.
Once some subset of ions has been deshelved to $\Sqb$, we drive the strong $\Sqb-\Pqb$ and ${5 ^{2}D_{3/2}}-\Pqb$ transitions with global beams at 493 and 650 nm, which either cools the ions or optically pumps them to $\ket{0_S}$.

We use this mechanism twice: first in (iii) to deshelve the $\ket{1_D}$ qubit state of the readout ancillae in order to perform the state readout, and then in (iv) to sympathetically cool the chain with the ancillae and reset them to $\ket{0_S}$.
Finally, in (v) and (vi), we reverse this mechanism to reinitialize the ancillae to $\ket{0_D}$ before deshelving the entire chain and resuming regular gate operations.

Detecting fluorescence during steps (iii) and (iv) provides two bits of information per MCM round. The first is the measurement bit, where a bright detection constitutes a projective measurement of $\ket{1_D}$. The second is the check bit, where unexpected fluorescence (i.e., a dark ancilla or bright non-ancilla qubit) indicates an ion that has leaked out of its proper qubit manifold. This leakage can be caused by either technical imperfections in the MCM sequence or spontaneous emission during Raman gates.
We do not convert detected leakage into erasures but rather post-select, rejecting shots in which leakage was detected during any MCM round or in a final check at the end of the circuit (see \cref{app:Postselect} for rejection rates).
As with a similar technique developed in neutral atoms~\cite{Thompson2023}, this reduces the MCM detection infidelity, idle errors of non-ancilla qubits during MCM, and one- and two-qubit gate errors.

We remark that this form of post-selection, which has also been used in other QEC demonstrations~\cite{wang_demonstration_2026, dasu2026computing}, can be converted to erasures with conditional reset of afflicted ions.
To justify the scalability of this approach and demonstrate that leakages do not significantly affect the logical error rate, we simulate re-initializing a leaked qubit rather than discarding the entire measurement (see~\cref{app:Postselect}). We find that in the regime of our experiments, doing so only degrades the logical error rate by about 10-17\%.

To our knowledge, our system constitutes the first demonstration of the full OMG architecture in a fully-functional trapped-ion quantum computer. Components of the architecture have been demonstrated previously in two ions \cite{Yu2025OMG,Yang2022OMG} or a single ion \cite{Chen2025OMG}, but this work is the first to implement it with both the scale, in terms of the number of qubits and the ability to entangle them, and the fidelity required to perform computations comparable to those demonstrated here.

\section{Quantum Memory experiments}

All the codes considered in this work are stabilizer codes~\cite{gottesman1997stabilizer}.
Recall that a {\em stabilizer code} with parameters $[[n,k,d]]$ is defined by a set of {\em stabilizer generators} $S_1, \dots, S_{n-k}$ that are independent commuting Pauli operators acting on $n$ qubits.
The code space is a $2^k$-dimensional subspace of the $n$-qubit Hilbert space, comprised of the set of {\em code states} $\ket{\psi}$ satisfying $S_i \ket \psi = \ket \psi$ for all $i$.
We interpret $\ket{\psi}$ as a state of $k$ {\em logical qubits} encoded within $n$ {\em physical qubits}.
The {\em minimum distance}, $d$, of the code measures its error-correction capability, with any error acting on up to $\lfloor (d-1)/2\rfloor$ physical qubits being correctable.
In this work, we implement only {\em CSS codes} where each $S_i$ is a product of $I$ and either $X$ or $Z$, but not both.

To protect a code state from error, a {\em syndrome circuit} is executed repeatedly, which measures stabilizer generators $S_i$ of the code.
In the absence of errors, these measurements return a trivial outcome since $\ket{\psi}$ is a +1-eigenstate of $S_i$.
Conversely, non-trivial measurement outcomes indicate presence of error.
We refer to one set of measurements of all stabilizer generators as a {\em syndrome extraction cycle}, and the corresponding duration as a {\em cycle time}.
Syndrome data gathered over one or more syndrome extraction cycles may be used to infer corrections, using a classical algorithm called the {\em decoder}.

A Pauli operator $P$, written as the product of single-qubit operators $P = P_1 P_2 \dots P_n$, can be measured by successively applying two-qubit controlled-$P_j$ gates for all $j$'s, controlled on a single $\ket{+}$ ancilla.
Measurement of the ancilla in the $X$ basis then yields the value of $P$.
Our syndrome circuits are constructed similarly, and we designate $n_a\leq 40-n$ out of forty ions as ancillae.
At any given time, each ancilla ion is assigned to measure a specific stabilizer generator $S_i$.
Controlled-Pauli gates corresponding to each $S_i$ are implemented sequentially.
For illustration, \Cref{fig:MainFig}B shows our syndrome circuit for the $[[4,2,2]]$ code using $n_a=2$.

On our system, the $n_a$ ancillae are all measured simultaneously, in consolidated MCM rounds (\cref{fig:MainFig}B ``MCM'' box, and \cref{fig:MainFig}D), then simultaneously reset and re-assigned to subsequent stabilizer generators.
Batched ancillae measurement following sequential gates was previously considered and optimized for small qLDPC codes~\cite{ye_quantum_2025}; here we chose $10\leq n_a\leq 16$, subject to ion availability.
We repeat the syndrome extraction cycle $\nSEC$ times in each experiment, and we pipeline ancilla usage such that ancillae that are read out in a single MCM round may measure stabilizer generators from consecutive syndrome extraction cycles.
This pipelining is intended to avoid idling unassigned ancillae and to minimize the number of MCM rounds to at most $\lceil \nSEC(n-k) / n_a\rceil$.
While all of our syndrome circuits follow the template of \cref{fig:MainFig}B, many permutations of stabilizer generators, and ordering of gates implementing their measurements, are possible.
For brevity, we refer interested readers to \cref{app:SyndromeCircuits}, where we detail specific gate and $S_i$ ordering, along with $n_a$ for each code.

\begin{figure*}[t]
\begin{centering}
\begin{tabular}{cc}
\tabularnewline
\includegraphics[width=8cm]{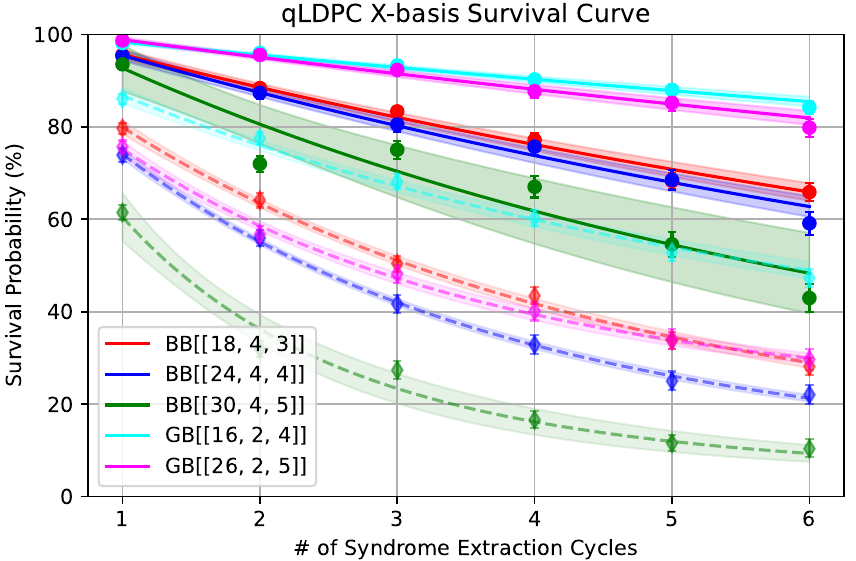}
\hspace{0.2in}
\includegraphics[width=8cm]{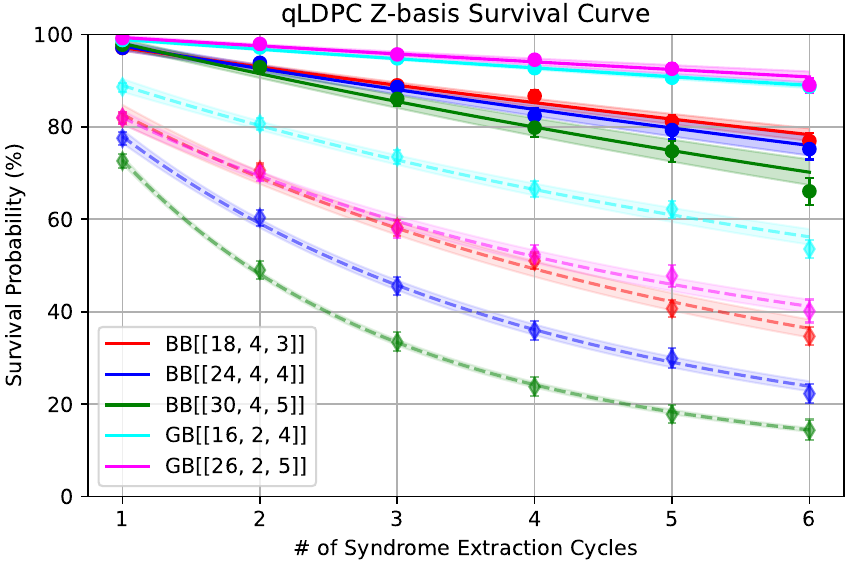}\tabularnewline
\tabularnewline
\end{tabular}
\par\end{centering}
\caption{
Survival probabilities $1-\plog$ for the BB5 and GB4 codes versus number of syndrome extraction cycles, for $X$ (left) and $Z$ (right) eigenstate memory experiments.
Circles and solid curves indicate the probability that no logical fault occurred in any of the encoded logical qubits.
Diamonds and dashed curves indicate the corresponding probabilities without decoding. Error bars around each point depict the 95\% Wilson confidence intervals for our estimate of $\plog$, and shaded bands show 95\% confidence bands around each fitted curve.
}
\label{fig:Survival}
\end{figure*}

Our experimental demonstrations encompass three families of quantum error correction codes: five distinct qLDPC codes, two topological codes on a toric manifold, and a concatenated code.
Our qLDPC codes consist of three BB5 codes and two GB4 codes (GB codes with weight-4 stabilizer generators) -- see \cref{app:CodePolynomials} for a full description of each code.
The topological codes are a $d=3$ standard toric code~\cite{bravyi_quantum_1998} and a $d=4$ rotated variant~\cite{bombin_optimal_2007,a_a_kovalev_improved_2012}.
The concatenated code is the $[[4,2,2]]$ code concatenated with itself.
The patterns of controlled-Pauli gates needed to implement the syndrome circuits vary greatly across codes, and especially across code families.
In order to accommodate just one qLDPC instance, Wang {\em et al.}, built bespoke ``air-bridge'' structures alongside Josephson junctions with carefully optimized resistances~\cite{wang_demonstration_2026,zhang_digital_2022}.
It is remarkable, therefore, that our all-to-all gate connectivity affords us the flexibility to implement all three families of codes with no hardware reconfiguration.
Furthermore, we leverage that flexibility to choose advantageous mappings of qubits to ions, which lets us increase the average fidelity of the two-qubit gates used by each circuit with no change to the underlying hardware, calibrations, or circuit structure (see \cref{app:Qubit Mapping}).

We performed {\em memory experiments}, where a quantum state is prepared and then kept coherent for as long as possible.
We repeat the syndrome extraction cycle, which corresponds to one full set of stabilizer generator measurements, several times.
For each code (encoding $k$ logical qubits), we prepare four logical states corresponding to $\pm 1$-eigenstates of $Z_L^k$ or $X_L^k$ (here, the subscript $L$ denotes a logical operator).
For $+1$-eigenstates, it suffices to prepare all physical qubits in $\ket{0}^n$ or $\ket{1}^n$; for $-1$-eigenstates, this is followed by one layer of single-qubit Pauli operators targeting physical qubits that implement $Z_L$ or $X_L$.
Each logical state preparation is subjected to a number of syndrome extraction cycles $\nSEC\in\{1,2,...,6\}$, after which all qubits of the code are destructively measured in $X$ or $Z$ respectively.
We run each memory experiment $2,000$ times, for a total of $48,000$ shots per code instance.
All syndrome data and qubit measurement outcomes are then decoded using the beam decoder with width 32 and 340 iterations~\cite{ye_beam_2025}.

\section{Results}
A logical error is said to occur if, upon decoding, any of the code's $k$ logical qubits are found to be in states at the end of a memory experiment that differ from their initial preparation.
For each code instance and state preparation basis, we calculate the {\em logical error rate per cycle} $\plog$ across syndrome extraction cycles $\nSEC$ and fit to:
\begin{align}
    1-\plog = A\left(\frac{1+e^{-\nSEC/\tau}}{2}\right)^{k}
    \label{eq:Fit}
\end{align}
with fit parameters $\tau$ and $A$.
From that fit, we infer the {\em logical error rate per cycle per qubit} as, $p_L = \frac{1}{2}(1-e^{-1/\tau})$.
\Cref{fig:Survival} depicts plots of $1-\plog$, with their fits as survival curves, for our qLDPC code implementations.
\Cref{tab:Survival} tabulates $p_L$ and $\tau$, showing the logical performance of the codes we implement.
We also show the BB6 $[[18,4,4]]$ code from Wang {\em et al.}~\cite{wang_demonstration_2026}.
Compared to our BB5 $[[18,4,3]]$ implementation, which contains the same number of physical and logical qubits, we achieve $4\times$ and $9\times$ lower logical error rates for $X_L$ and $Z_L$ eigenstate memory experiments, which are respectively sensitive to phase- and bit-flip errors.

\begin{table}
\begin{centering}
\begin{tabular}{|c|c|c|c|c|}
\hline 
 & \multicolumn{2}{c|}{$X$ eigenstate} & \multicolumn{2}{c|}{$Z$ eigenstate}\tabularnewline
\hline 
Code & $\tau$ & $p_{L}$ ($10^{-2}$) & $\tau$ & $p_{L}$ ($10^{-2}$)\tabularnewline
\hline 
\hline 
BB$[[18,4,3]]$  & $25.1\pm4.31$ & $1.95\pm0.39$ & $44.9\pm5.45$ & $1.10\pm0.15$ \tabularnewline
\hline 
BB$[[24,4,4]]$  & $22.4\pm3.01$ & $2.19\pm0.33$ & $38.4\pm6.01$ & $1.28\pm0.23$ \tabularnewline
\hline 
BB$[[30,4,5]]$  & $13.7\pm6.59$ & $3.52\pm3.05$ & $29.0\pm5.80$ & $1.70\pm0.42$ \tabularnewline
\hline 
GB$[[16,2,4]]$  & $34.4\pm5.30$ & $1.43\pm0.26$ & $47.0\pm3.89$ & $1.05\pm0.09$ \tabularnewline
\hline 
GB$[[26,2,5]]$  & $25.6\pm4.72$ & $1.91\pm0.42$ & $54.7\pm8.84$ & $0.91\pm0.17$ \tabularnewline
\hline 
Tor$[[18,2,3]]$ & $34.0\pm6.82$ & $1.45\pm0.36$ & $65.5\pm23.00$ & $0.76\pm0.40$ \tabularnewline
\hline 
Tor$[[16,2,4]]$ & $35.1\pm4.03$ & $1.40\pm0.18$ & $51.0\pm18.14$ & $0.97\pm0.53$ \tabularnewline
\hline 
Con$[[16,4,4]]$ & $18.6\pm3.13$ & $2.61\pm0.51$ & $23.6\pm3.22$ & $2.07\pm0.32$ \tabularnewline
\hline 
$[[18,4,4]]$~\cite{wang_demonstration_2026} & -  & $8.67\pm0.21$ & - & $9.15\pm0.27$\tabularnewline
  \hline 
  \end{tabular}
  \par\end{centering}
  \caption{Logical error rate per logical qubit per syndrome extraction cycle $p_L$, and fit $\tau$, for memory experiments with $X$ and $Z$ eigenstates.
  ``Tor''=Toric and ``Con''=Concatenated.
  The BB6 code of~\cite{wang_demonstration_2026} is also shown for comparison.}
  \label{tab:Survival}
\end{table}

To facilitate comparisons between physical and logical qubits, we multiply $\tau$ for each code by its cycle time (see \cref{app:Syndrome Circuit Timing}) to obtain $1/e$ survival times $T_X$ and $T_Z$ for the $X_L$ and $Z_L$ eigenstates.
We remark that, by construction, $T_X$ measures survival time of an $X_L$ eigenstate in exactly the same sense as $\Ttwo$ relaxations do for physical qubits under white phase noise, as measured by standard Ramsey experiments.
Our qubits were measured to have $\Ttwo\approx 1.28\pm 0.16$ s (\cref{app:T2star}).

For a more general comparison of qubit memories, we turn to a notion of lifetime similar to that introduced in the below-breakeven surface code experiment by Google Quantum AI {\em et al.}~\cite{acharya_quantum_2025}.
Let $\Noiset$ be a quantum channel representing the noise accumulated on a qubit over a time $t \geq 0$.
Define the {\em channel decay speed} as
\begin{align}
    D_\Noiset := - \frac{d}{dt} F_{\Noise}(t)
\end{align}
where $F_{\Noise}(t)$ is the process fidelity of $\Noiset$ with respect to the trivial identity channel~\cite{pedersen_fidelity_2007}.
The channel decay speed measures the rapidity at which the fidelity drops.
We define the {\em lifetime} of a qubit that undergoes the channel $\Noiset$, to be $L_\Noise := 1/D_\Noise(0)$. This generalizes the notion of a $1/e$ survival time, under exponential decay, to be independent of specific preparation bases.
We take $\Noiset$ to be decay channels natural to each type of qubit.
Our definition of lifetime differs from~\cite{acharya_quantum_2025} only in the constraint imposed while inferring the parameters of $\Noiset$ (see \cref{app:Physical Logical Qubit Comparison}).

\Cref{tab:SurvivalRealtime} summarizes lifetimes for all the codes we implement.
We remark that, in all codes except the concatenated $[[4,2,2]]$, the logical qubit lifetimes are comparable to our physical qubits' to within error bars, and they even exceed it slightly in several cases.

\begin{table}
  \begin{centering}
\begin{tabular}{|c|c|c|c|}
\hline 
Code & $T_{X}$ (s) & $T_{Z}$ (s) & $T_{\Noise}$ (s)\tabularnewline
\hline 
\hline 
BB$[[18,4,3]]$  & $1.06\pm0.18$ & $1.90\pm0.23$ & $2.48\pm0.40$ \tabularnewline
\hline 
BB$[[24,4,4]]$  & $1.46\pm0.20$ & $2.51\pm0.39$ & $3.39\pm0.47$ \tabularnewline
\hline 
BB$[[30,4,5]]$  & $1.17\pm0.57$ & $2.48\pm0.50$ & $2.85\pm1.26$ \tabularnewline
\hline 
GB$[[16,2,4]]$  & $1.21\pm0.19$ & $1.65\pm0.14$ & $2.65\pm0.36$ \tabularnewline
\hline 
GB$[[26,2,5]]$  & $1.63\pm0.30$ & $3.47\pm0.56$ & $3.95\pm0.71$ \tabularnewline
\hline 
Tor$[[18,2,3]]$ & $1.33\pm0.27$ & $2.57\pm0.90$ & $3.17\pm0.75$ \tabularnewline
\hline 
Tor$[[16,2,4]]$ & $1.23\pm0.14$ & $1.79\pm0.64$ & $2.75\pm0.53$ \tabularnewline
\hline 
Con$[[16,4,4]]$ & $0.72\pm0.12$ & $0.92\pm0.12$ & $1.55\pm0.25$ \tabularnewline
\hline 
Physical &  $1.28\pm0.16$ & $\infty$ & $3.84\pm0.48$\tabularnewline
\hline 
\end{tabular}
  \par\end{centering}
  \caption{Logical qubit lifetimes in seconds.
  The $1/e$ survival times $T_X$ \& $T_Z$ for eigenstates of $X_L$ \& $Z_L$ are shown, as well as the overall lifetime $T_{\Noise}$ of each qubit undergoing decay under channel $\Noise$.
  For physical qubits, values shown are $\Ttwo$ and $\Tone$ respectively.}
  \label{tab:SurvivalRealtime}
\end{table}

\section{Conclusion}

We described several quantum error correction experiments in the breakeven regime, demonstrating three families of codes that have played a critical role in the development of the theory of quantum error correction and which today have the potential to enable the first generations of fault-tolerant quantum computers.
With a BB code, we demonstrate logical error rates $4\times$ and $9\times$ lower for phase- and bit-flip errors compared to a prior experiment on superconducting qubits~\cite{wang_demonstration_2026}.
The system on which this work was demonstrated occupies an architectural sweet spot, being large enough to accommodate non-trivial qLDPC code blocks while small enough to retain performant all-to-all physical gates without shuttling.
Our OMG implementation -- a first in a fully-functioning trapped-ion quantum computer -- bolsters that flexibility.

As we transition toward larger systems, it would be beneficial to increase parallelism to reduce the execution time of large quantum circuits.
Increasing the number of Raman zones, for example, speeds up gates and selective shelving/deshelving operations.
A more scalable direction is based on electronic qubit control~\cite{malinowski2023wire}, increasing gate parallelism while leveraging record-fidelity gates~\cite{loschnauer2025scalable, hughes2025trapped, walking_cat}.
Furthermore enhancements like dynamical decoupling, which we do not use here, can significantly improve both the $\Ttwo$ of physical qubits and the lifetimes of logical qubits.

A perennial challenge in implementing QEC lies in reconciling hardware constraints of specific systems with the needs of particular codes.
Quantum LDPC codes, which are a leading contender for qubit-efficient fault-tolerant architectures thanks to their high encoding-rate, are notoriously challenging to implement because of their two-qubit gate connectivity requirements.
Trapped ions, which boast record-fidelity gates and a high degree of flexibility and reconfigurability, are a natural platform for the implementation of qLDPC codes.
This work highlights the implementation of high-rate qLDPC codes leveraging the flexibility of a trapped-ion device, and therefore represents an important milestone towards a qubit-efficient fault-tolerant architecture.

\section*{Acknowledgments}
The authors acknowledge the broader IonQ team, in particular Chris Ballance and David Allcock for facilitating the project.
The authors also acknowledge Max Ballenger and Yash Bansood for sharing their insights on the software stack used in this experiment.

\section*{Author Contributions}
ET and MLG jointly led the experimental effort.
ET led the QEC experiment design, optimization, and data analyses.
MLG led the development of the OMG architecture and oversaw execution of the experiment on our system.
SD and ANP contributed substantially to the regular operation of the experimental system and to data acquisition.
MB, JM, JN, JS, and AW assisted in developing and building the experimental system.
EN led the QCVV and system characterisation efforts.
MR provided crucial software support.
FT contributed substantially to the search and construction of new high-rate qLDPC codes we demonstrated.
MY built the decoder we used.
NP, KW, and JG provided project direction.
ND provided critical technical insight and overall project direction.

\newpage

\pagebreak
\bibliography{references}

\begin{thebibliography}{55}%
\makeatletter
\providecommand \@ifxundefined [1]{%
 \@ifx{#1\undefined}
}%
\providecommand \@ifnum [1]{%
 \ifnum #1\expandafter \@firstoftwo
 \else \expandafter \@secondoftwo
 \fi
}%
\providecommand \@ifx [1]{%
 \ifx #1\expandafter \@firstoftwo
 \else \expandafter \@secondoftwo
 \fi
}%
\providecommand \natexlab [1]{#1}%
\providecommand \enquote  [1]{``#1''}%
\providecommand \bibnamefont  [1]{#1}%
\providecommand \bibfnamefont [1]{#1}%
\providecommand \citenamefont [1]{#1}%
\providecommand \href@noop [0]{\@secondoftwo}%
\providecommand \href [0]{\begingroup \@sanitize@url \@href}%
\providecommand \@href[1]{\@@startlink{#1}\@@href}%
\providecommand \@@href[1]{\endgroup#1\@@endlink}%
\providecommand \@sanitize@url [0]{\catcode `\\12\catcode `\$12\catcode `\&12\catcode `\#12\catcode `\^12\catcode `\_12\catcode `\%12\relax}%
\providecommand \@@startlink[1]{}%
\providecommand \@@endlink[0]{}%
\providecommand \url  [0]{\begingroup\@sanitize@url \@url }%
\providecommand \@url [1]{\endgroup\@href {#1}{\urlprefix }}%
\providecommand \urlprefix  [0]{URL }%
\providecommand \Eprint [0]{\href }%
\providecommand \doibase [0]{https://doi.org/}%
\providecommand \selectlanguage [0]{\@gobble}%
\providecommand \bibinfo  [0]{\@secondoftwo}%
\providecommand \bibfield  [0]{\@secondoftwo}%
\providecommand \translation [1]{[#1]}%
\providecommand \BibitemOpen [0]{}%
\providecommand \bibitemStop [0]{}%
\providecommand \bibitemNoStop [0]{.\EOS\space}%
\providecommand \EOS [0]{\spacefactor3000\relax}%
\providecommand \BibitemShut  [1]{\csname bibitem#1\endcsname}%
\let\auto@bib@innerbib\@empty
\bibitem [{\citenamefont {Shor}(1996)}]{shor1996fault}%
  \BibitemOpen
  \bibfield  {author} {\bibinfo {author} {\bibfnamefont {P.~W.}\ \bibnamefont {Shor}},\ }\bibfield  {title} {\bibinfo {title} {Fault-tolerant quantum computation},\ }in\ \href@noop {} {\emph {\bibinfo {booktitle} {Proceedings of 37th conference on foundations of computer science}}}\ (\bibinfo {organization} {IEEE},\ \bibinfo {year} {1996})\ pp.\ \bibinfo {pages} {56--65}\BibitemShut {NoStop}%
\bibitem [{\citenamefont {Dennis}\ \emph {et~al.}(2002)\citenamefont {Dennis}, \citenamefont {Kitaev}, \citenamefont {Landahl},\ and\ \citenamefont {Preskill}}]{dennis_topological_2002}%
  \BibitemOpen
  \bibfield  {author} {\bibinfo {author} {\bibfnamefont {E.}~\bibnamefont {Dennis}}, \bibinfo {author} {\bibfnamefont {A.}~\bibnamefont {Kitaev}}, \bibinfo {author} {\bibfnamefont {A.}~\bibnamefont {Landahl}},\ and\ \bibinfo {author} {\bibfnamefont {J.}~\bibnamefont {Preskill}},\ }\bibfield  {title} {\bibinfo {title} {Topological quantum memory},\ }\href {https://doi.org/10.1063/1.1499754} {\bibfield  {journal} {\bibinfo  {journal} {Journal of Mathematical Physics}\ }\textbf {\bibinfo {volume} {43}},\ \bibinfo {pages} {4452} (\bibinfo {year} {2002})}\BibitemShut {NoStop}%
\bibitem [{\citenamefont {Fowler}\ \emph {et~al.}(2012)\citenamefont {Fowler}, \citenamefont {Mariantoni}, \citenamefont {Martinis},\ and\ \citenamefont {Cleland}}]{fowler_surface_2012}%
  \BibitemOpen
  \bibfield  {author} {\bibinfo {author} {\bibfnamefont {A.~G.}\ \bibnamefont {Fowler}}, \bibinfo {author} {\bibfnamefont {M.}~\bibnamefont {Mariantoni}}, \bibinfo {author} {\bibfnamefont {J.~M.}\ \bibnamefont {Martinis}},\ and\ \bibinfo {author} {\bibfnamefont {A.~N.}\ \bibnamefont {Cleland}},\ }\bibfield  {title} {\bibinfo {title} {Surface codes: {Towards} practical large-scale quantum computation},\ }\href {https://doi.org/10.1103/PhysRevA.86.032324} {\bibfield  {journal} {\bibinfo  {journal} {Physical Review A}\ }\textbf {\bibinfo {volume} {86}},\ \bibinfo {pages} {032324} (\bibinfo {year} {2012})}\BibitemShut {NoStop}%
\bibitem [{\citenamefont {Krinner}\ \emph {et~al.}(2022)\citenamefont {Krinner} \emph {et~al.}}]{krinner2022realizing}%
  \BibitemOpen
  \bibfield  {author} {\bibinfo {author} {\bibfnamefont {S.}~\bibnamefont {Krinner}} \emph {et~al.},\ }\bibfield  {title} {\bibinfo {title} {Realizing repeated quantum error correction in a distance-three surface code},\ }\href@noop {} {\bibfield  {journal} {\bibinfo  {journal} {Nature}\ }\textbf {\bibinfo {volume} {605}},\ \bibinfo {pages} {669} (\bibinfo {year} {2022})}\BibitemShut {NoStop}%
\bibitem [{\citenamefont {Zhao}\ \emph {et~al.}(2022)\citenamefont {Zhao} \emph {et~al.}}]{zhao2022realization}%
  \BibitemOpen
  \bibfield  {author} {\bibinfo {author} {\bibfnamefont {Y.}~\bibnamefont {Zhao}} \emph {et~al.},\ }\bibfield  {title} {\bibinfo {title} {Realization of an error-correcting surface code with superconducting qubits},\ }\href@noop {} {\bibfield  {journal} {\bibinfo  {journal} {Physical Review Letters}\ }\textbf {\bibinfo {volume} {129}},\ \bibinfo {pages} {030501} (\bibinfo {year} {2022})}\BibitemShut {NoStop}%
\bibitem [{\citenamefont {Bluvstein}\ \emph {et~al.}(2024)\citenamefont {Bluvstein} \emph {et~al.}}]{bluvstein_logical_2024}%
  \BibitemOpen
  \bibfield  {author} {\bibinfo {author} {\bibfnamefont {D.}~\bibnamefont {Bluvstein}} \emph {et~al.},\ }\bibfield  {title} {\bibinfo {title} {Logical quantum processor based on reconfigurable atom arrays},\ }\href {https://doi.org/10.1038/s41586-023-06927-3} {\bibfield  {journal} {\bibinfo  {journal} {Nature}\ }\textbf {\bibinfo {volume} {626}},\ \bibinfo {pages} {58} (\bibinfo {year} {2024})}\BibitemShut {NoStop}%
\bibitem [{\citenamefont {{Google Quantum AI}}(2025)}]{acharya_quantum_2025}%
  \BibitemOpen
  \bibfield  {author} {\bibinfo {author} {\bibnamefont {{Google Quantum AI}}},\ }\bibfield  {title} {\bibinfo {title} {Quantum error correction below the surface code threshold},\ }\href {https://doi.org/10.1038/s41586-024-08449-y} {\bibfield  {journal} {\bibinfo  {journal} {Nature}\ }\textbf {\bibinfo {volume} {638}},\ \bibinfo {pages} {920} (\bibinfo {year} {2025})}\BibitemShut {NoStop}%
\bibitem [{\citenamefont {He}\ \emph {et~al.}(2025)\citenamefont {He} \emph {et~al.}}]{he2025experimental}%
  \BibitemOpen
  \bibfield  {author} {\bibinfo {author} {\bibfnamefont {T.}~\bibnamefont {He}} \emph {et~al.},\ }\bibfield  {title} {\bibinfo {title} {Experimental quantum error correction below the surface code threshold via all-microwave leakage suppression},\ }\href@noop {} {\bibfield  {journal} {\bibinfo  {journal} {Physical Review Letters}\ }\textbf {\bibinfo {volume} {135}},\ \bibinfo {pages} {260601} (\bibinfo {year} {2025})}\BibitemShut {NoStop}%
\bibitem [{\citenamefont {MacKay}\ \emph {et~al.}(2004)\citenamefont {MacKay}, \citenamefont {Mitchison},\ and\ \citenamefont {McFadden}}]{mackay2004sparse}%
  \BibitemOpen
  \bibfield  {author} {\bibinfo {author} {\bibfnamefont {D.~J.}\ \bibnamefont {MacKay}}, \bibinfo {author} {\bibfnamefont {G.}~\bibnamefont {Mitchison}},\ and\ \bibinfo {author} {\bibfnamefont {P.~L.}\ \bibnamefont {McFadden}},\ }\bibfield  {title} {\bibinfo {title} {Sparse-graph codes for quantum error correction},\ }\href@noop {} {\bibfield  {journal} {\bibinfo  {journal} {IEEE Transactions on Information Theory}\ }\textbf {\bibinfo {volume} {50}},\ \bibinfo {pages} {2315} (\bibinfo {year} {2004})}\BibitemShut {NoStop}%
\bibitem [{\citenamefont {Breuckmann}\ and\ \citenamefont {Eberhardt}(2021)}]{breuckmann2021quantum}%
  \BibitemOpen
  \bibfield  {author} {\bibinfo {author} {\bibfnamefont {N.~P.}\ \bibnamefont {Breuckmann}}\ and\ \bibinfo {author} {\bibfnamefont {J.~N.}\ \bibnamefont {Eberhardt}},\ }\bibfield  {title} {\bibinfo {title} {Quantum low-density parity-check codes},\ }\href@noop {} {\bibfield  {journal} {\bibinfo  {journal} {PRX quantum}\ }\textbf {\bibinfo {volume} {2}},\ \bibinfo {pages} {040101} (\bibinfo {year} {2021})}\BibitemShut {NoStop}%
\bibitem [{\citenamefont {Xu}\ \emph {et~al.}(2024)\citenamefont {Xu} \emph {et~al.}}]{xu2024constant}%
  \BibitemOpen
  \bibfield  {author} {\bibinfo {author} {\bibfnamefont {Q.}~\bibnamefont {Xu}} \emph {et~al.},\ }\bibfield  {title} {\bibinfo {title} {Constant-overhead fault-tolerant quantum computation with reconfigurable atom arrays},\ }\href@noop {} {\bibfield  {journal} {\bibinfo  {journal} {Nature Physics}\ }\textbf {\bibinfo {volume} {20}},\ \bibinfo {pages} {1084} (\bibinfo {year} {2024})}\BibitemShut {NoStop}%
\bibitem [{\citenamefont {Yoder}\ \emph {et~al.}(2025)\citenamefont {Yoder} \emph {et~al.}}]{yoder2025tour}%
  \BibitemOpen
  \bibfield  {author} {\bibinfo {author} {\bibfnamefont {T.~J.}\ \bibnamefont {Yoder}} \emph {et~al.},\ }\bibfield  {title} {\bibinfo {title} {Tour de gross: A modular quantum computer based on bivariate bicycle codes},\ }\href@noop {} {\bibfield  {journal} {\bibinfo  {journal} {{P}reprint at https://arxiv.org/abs/2506.03094}\ } (\bibinfo {year} {2025})}\BibitemShut {NoStop}%
\bibitem [{\citenamefont {Webster}\ \emph {et~al.}(2026)\citenamefont {Webster} \emph {et~al.}}]{webster2026pinnacle}%
  \BibitemOpen
  \bibfield  {author} {\bibinfo {author} {\bibfnamefont {P.}~\bibnamefont {Webster}} \emph {et~al.},\ }\bibfield  {title} {\bibinfo {title} {The pinnacle architecture: Reducing the cost of breaking rsa-2048 to 100 000 physical qubits using quantum ldpc codes},\ }\href@noop {} {\bibfield  {journal} {\bibinfo  {journal} {{P}reprint at https://arxiv.org/abs/2602.11457}\ } (\bibinfo {year} {2026})}\BibitemShut {NoStop}%
\bibitem [{\citenamefont {Cain}\ \emph {et~al.}(2026)\citenamefont {Cain} \emph {et~al.}}]{cain2026shor}%
  \BibitemOpen
  \bibfield  {author} {\bibinfo {author} {\bibfnamefont {M.}~\bibnamefont {Cain}} \emph {et~al.},\ }\bibfield  {title} {\bibinfo {title} {Shor's algorithm is possible with as few as 10,000 reconfigurable atomic qubits},\ }\href@noop {} {\bibfield  {journal} {\bibinfo  {journal} {{P}reprint at https://arxiv.org/abs/2603.28627}\ } (\bibinfo {year} {2026})}\BibitemShut {NoStop}%
\bibitem [{\citenamefont {Tripier}\ \emph {et~al.}(2026{\natexlab{a}})\citenamefont {Tripier} \emph {et~al.}}]{tripier2026fault}%
  \BibitemOpen
  \bibfield  {author} {\bibinfo {author} {\bibfnamefont {F.}~\bibnamefont {Tripier}} \emph {et~al.},\ }\bibfield  {title} {\bibinfo {title} {Fault-tolerant quantum computing with trapped ions: The walking cat architecture},\ }\href@noop {} {\bibfield  {journal} {\bibinfo  {journal} {{P}reprint at https://arxiv.org/abs/2604.19481}\ } (\bibinfo {year} {2026}{\natexlab{a}})}\BibitemShut {NoStop}%
\bibitem [{\citenamefont {Tremblay}\ \emph {et~al.}(2022)\citenamefont {Tremblay}, \citenamefont {Delfosse},\ and\ \citenamefont {Beverland}}]{tremblay2022constant}%
  \BibitemOpen
  \bibfield  {author} {\bibinfo {author} {\bibfnamefont {M.~A.}\ \bibnamefont {Tremblay}}, \bibinfo {author} {\bibfnamefont {N.}~\bibnamefont {Delfosse}},\ and\ \bibinfo {author} {\bibfnamefont {M.~E.}\ \bibnamefont {Beverland}},\ }\bibfield  {title} {\bibinfo {title} {Constant-overhead quantum error correction with thin planar connectivity},\ }\href@noop {} {\bibfield  {journal} {\bibinfo  {journal} {Physical Review Letters}\ }\textbf {\bibinfo {volume} {129}},\ \bibinfo {pages} {050504} (\bibinfo {year} {2022})}\BibitemShut {NoStop}%
\bibitem [{\citenamefont {Aydin}\ \emph {et~al.}(2026)\citenamefont {Aydin}, \citenamefont {Delfosse},\ and\ \citenamefont {Tham}}]{aydin_cyclic_2026}%
  \BibitemOpen
  \bibfield  {author} {\bibinfo {author} {\bibfnamefont {A.}~\bibnamefont {Aydin}}, \bibinfo {author} {\bibfnamefont {N.}~\bibnamefont {Delfosse}},\ and\ \bibinfo {author} {\bibfnamefont {E.}~\bibnamefont {Tham}},\ }\bibfield  {title} {\bibinfo {title} {Cyclic {Hypergraph} {Product} {Code}},\ }\href@noop {} {\bibfield  {journal} {\bibinfo  {journal} {{P}reprint at https://doi.org/10.48550/arXiv.2511.09683}\ } (\bibinfo {year} {2026})}\BibitemShut {NoStop}%
\bibitem [{\citenamefont {Bravyi}\ \emph {et~al.}(2024)\citenamefont {Bravyi} \emph {et~al.}}]{bravyi_high-threshold_2024}%
  \BibitemOpen
  \bibfield  {author} {\bibinfo {author} {\bibfnamefont {S.}~\bibnamefont {Bravyi}} \emph {et~al.},\ }\bibfield  {title} {\bibinfo {title} {High-threshold and low-overhead fault-tolerant quantum memory},\ }\href {http://dx.doi.org/10.1038/s41586-024-07107-7} {\bibfield  {journal} {\bibinfo  {journal} {Nature}\ }\textbf {\bibinfo {volume} {627}},\ \bibinfo {pages} {778} (\bibinfo {year} {2024})}\BibitemShut {NoStop}%
\bibitem [{\citenamefont {Ye}\ and\ \citenamefont {Delfosse}(2025)}]{ye_quantum_2025}%
  \BibitemOpen
  \bibfield  {author} {\bibinfo {author} {\bibfnamefont {M.}~\bibnamefont {Ye}}\ and\ \bibinfo {author} {\bibfnamefont {N.}~\bibnamefont {Delfosse}},\ }\bibfield  {title} {\bibinfo {title} {Quantum error correction for long chains of trapped ions},\ }\href@noop {} {\bibfield  {journal} {\bibinfo  {journal} {{P}reprint at https://doi.org/10.48550/arXiv.2503.22071}\ } (\bibinfo {year} {2025})}\BibitemShut {NoStop}%
\bibitem [{\citenamefont {Scruby}\ \emph {et~al.}(2026)\citenamefont {Scruby}, \citenamefont {Hillmann},\ and\ \citenamefont {Roffe}}]{scruby2026high}%
  \BibitemOpen
  \bibfield  {author} {\bibinfo {author} {\bibfnamefont {T.~R.}\ \bibnamefont {Scruby}}, \bibinfo {author} {\bibfnamefont {T.}~\bibnamefont {Hillmann}},\ and\ \bibinfo {author} {\bibfnamefont {J.}~\bibnamefont {Roffe}},\ }\bibfield  {title} {\bibinfo {title} {High-threshold, low-overhead and single-shot decodable fault-tolerant quantum memory},\ }\href@noop {} {\bibfield  {journal} {\bibinfo  {journal} {PRX Quantum}\ }\textbf {\bibinfo {volume} {7}},\ \bibinfo {pages} {020310} (\bibinfo {year} {2026})}\BibitemShut {NoStop}%
\bibitem [{\citenamefont {Wang}\ \emph {et~al.}(2026)\citenamefont {Wang} \emph {et~al.}}]{wang_demonstration_2026}%
  \BibitemOpen
  \bibfield  {author} {\bibinfo {author} {\bibfnamefont {K.}~\bibnamefont {Wang}} \emph {et~al.},\ }\bibfield  {title} {\bibinfo {title} {Demonstration of low-overhead quantum error correction codes},\ }\href {https://doi.org/10.1038/s41567-025-03157-4} {\bibfield  {journal} {\bibinfo  {journal} {Nature Physics}\ } (\bibinfo {year} {2026})}\BibitemShut {NoStop}%
\bibitem [{\citenamefont {Ryan-Anderson}\ \emph {et~al.}(2022)\citenamefont {Ryan-Anderson} \emph {et~al.}}]{ryan-anderson_implementing_2022}%
  \BibitemOpen
  \bibfield  {author} {\bibinfo {author} {\bibfnamefont {C.}~\bibnamefont {Ryan-Anderson}} \emph {et~al.},\ }\bibfield  {title} {\bibinfo {title} {Implementing {Fault}-tolerant {Entangling} {Gates} on the {Five}-qubit {Code} and the {Color} {Code}},\ }\href@noop {} {\bibfield  {journal} {\bibinfo  {journal} {{P}reprint at https://doi.org/10.48550/arXiv.2208.01863}\ } (\bibinfo {year} {2022})}\BibitemShut {NoStop}%
\bibitem [{\citenamefont {Egan}\ \emph {et~al.}(2021)\citenamefont {Egan} \emph {et~al.}}]{egan_fault_tolerant_2021}%
  \BibitemOpen
  \bibfield  {author} {\bibinfo {author} {\bibfnamefont {L.}~\bibnamefont {Egan}} \emph {et~al.},\ }\bibfield  {title} {\bibinfo {title} {Fault-tolerant control of an error-corrected qubit},\ }\href {https://doi.org/10.1038/s41586-021-03928-y} {\bibfield  {journal} {\bibinfo  {journal} {Nature}\ }\textbf {\bibinfo {volume} {598}},\ \bibinfo {pages} {281} (\bibinfo {year} {2021})}\BibitemShut {NoStop}%
\bibitem [{\citenamefont {Ryan-Anderson}\ \emph {et~al.}(2024)\citenamefont {Ryan-Anderson} \emph {et~al.}}]{ryan2024high}%
  \BibitemOpen
  \bibfield  {author} {\bibinfo {author} {\bibfnamefont {C.}~\bibnamefont {Ryan-Anderson}} \emph {et~al.},\ }\bibfield  {title} {\bibinfo {title} {High-fidelity teleportation of a logical qubit using transversal gates and lattice surgery},\ }\href@noop {} {\bibfield  {journal} {\bibinfo  {journal} {Science}\ }\textbf {\bibinfo {volume} {385}},\ \bibinfo {pages} {1327} (\bibinfo {year} {2024})}\BibitemShut {NoStop}%
\bibitem [{\citenamefont {Paetznick}\ \emph {et~al.}(2024)\citenamefont {Paetznick} \emph {et~al.}}]{paetznick_demonstration_2024}%
  \BibitemOpen
  \bibfield  {author} {\bibinfo {author} {\bibfnamefont {A.}~\bibnamefont {Paetznick}} \emph {et~al.},\ }\bibfield  {title} {\bibinfo {title} {Demonstration of logical qubits and repeated error correction with better-than-physical error rates},\ }\href@noop {} {\bibfield  {journal} {\bibinfo  {journal} {{P}reprint at https://doi.org/10.48550/arXiv.2404.02280}\ } (\bibinfo {year} {2024})}\BibitemShut {NoStop}%
\bibitem [{\citenamefont {Reichardt}\ \emph {et~al.}(2024)\citenamefont {Reichardt} \emph {et~al.}}]{reichardt2024demonstration}%
  \BibitemOpen
  \bibfield  {author} {\bibinfo {author} {\bibfnamefont {B.~W.}\ \bibnamefont {Reichardt}} \emph {et~al.},\ }\bibfield  {title} {\bibinfo {title} {Demonstration of quantum computation and error correction with a tesseract code},\ }\href@noop {} {\bibfield  {journal} {\bibinfo  {journal} {{P}reprint at https://arxiv.org/abs/2409.04628}\ } (\bibinfo {year} {2024})}\BibitemShut {NoStop}%
\bibitem [{\citenamefont {Aasen}\ \emph {et~al.}(2026)\citenamefont {Aasen} \emph {et~al.}}]{aasen2026}%
  \BibitemOpen
  \bibfield  {author} {\bibinfo {author} {\bibfnamefont {D.}~\bibnamefont {Aasen}} \emph {et~al.},\ }\bibfield  {title} {\bibinfo {title} {Quantum error correction with the toric code},\ }\href@noop {} {\bibfield  {journal} {\bibinfo  {journal} {Atom Computing}\ } (\bibinfo {year} {2026})}\BibitemShut {NoStop}%
\bibitem [{\citenamefont {Rines}\ \emph {et~al.}(2025)\citenamefont {Rines} \emph {et~al.}}]{rines2025demonstration}%
  \BibitemOpen
  \bibfield  {author} {\bibinfo {author} {\bibfnamefont {R.}~\bibnamefont {Rines}} \emph {et~al.},\ }\bibfield  {title} {\bibinfo {title} {Demonstration of a logical architecture uniting motion and in-place entanglement: Shor's algorithm, constant-depth cnot ladder, and many-hypercube code},\ }\href@noop {} {\bibfield  {journal} {\bibinfo  {journal} {{P}reprint at https://arxiv.org/abs/2509.13247}\ } (\bibinfo {year} {2025})}\BibitemShut {NoStop}%
\bibitem [{\citenamefont {Dasu}\ \emph {et~al.}(2026)\citenamefont {Dasu} \emph {et~al.}}]{dasu2026computing}%
  \BibitemOpen
  \bibfield  {author} {\bibinfo {author} {\bibfnamefont {S.}~\bibnamefont {Dasu}} \emph {et~al.},\ }\bibfield  {title} {\bibinfo {title} {Computing with many encoded logical qubits beyond break-even},\ }\href@noop {} {\bibfield  {journal} {\bibinfo  {journal} {{P}reprint at https://arxiv.org/abs/2602.22211}\ } (\bibinfo {year} {2026})}\BibitemShut {NoStop}%
\bibitem [{\citenamefont {Chen}\ \emph {et~al.}(2024)\citenamefont {Chen} \emph {et~al.}}]{Chen2024benchmarkingtrapped}%
  \BibitemOpen
  \bibfield  {author} {\bibinfo {author} {\bibfnamefont {J.-S.}\ \bibnamefont {Chen}} \emph {et~al.},\ }\bibfield  {title} {\bibinfo {title} {Benchmarking a trapped-ion quantum computer with 30 qubits},\ }\href {https://doi.org/10.22331/q-2024-11-07-1516} {\bibfield  {journal} {\bibinfo  {journal} {{Quantum}}\ }\textbf {\bibinfo {volume} {8}},\ \bibinfo {pages} {1516} (\bibinfo {year} {2024})}\BibitemShut {NoStop}%
\bibitem [{\citenamefont {Allcock}\ \emph {et~al.}(2021)\citenamefont {Allcock} \emph {et~al.}}]{allcock_omg_2021}%
  \BibitemOpen
  \bibfield  {author} {\bibinfo {author} {\bibfnamefont {D.~T.~C.}\ \bibnamefont {Allcock}} \emph {et~al.},\ }\bibfield  {title} {\bibinfo {title} {omg blueprint for trapped ion quantum computing with metastable states},\ }\href {https://doi.org/10.1063/5.0069544} {\bibfield  {journal} {\bibinfo  {journal} {Applied Physics Letters}\ }\textbf {\bibinfo {volume} {119}},\ \bibinfo {pages} {214002} (\bibinfo {year} {2021})}\BibitemShut {NoStop}%
\bibitem [{\citenamefont {Ransford}\ \emph {et~al.}(2025)\citenamefont {Ransford} \emph {et~al.}}]{ransford2025helios}%
  \BibitemOpen
  \bibfield  {author} {\bibinfo {author} {\bibfnamefont {A.}~\bibnamefont {Ransford}} \emph {et~al.},\ }\bibfield  {title} {\bibinfo {title} {Helios: A 98-qubit trapped-ion quantum computer},\ }\href@noop {} {\bibfield  {journal} {\bibinfo  {journal} {{P}reprint at https://arxiv.org/abs/2511.05465}\ } (\bibinfo {year} {2025})}\BibitemShut {NoStop}%
\bibitem [{\citenamefont {Moses}\ \emph {et~al.}(2023)\citenamefont {Moses} \emph {et~al.}}]{moses2023race}%
  \BibitemOpen
  \bibfield  {author} {\bibinfo {author} {\bibfnamefont {S.~A.}\ \bibnamefont {Moses}} \emph {et~al.},\ }\bibfield  {title} {\bibinfo {title} {A race-track trapped-ion quantum processor},\ }\href@noop {} {\bibfield  {journal} {\bibinfo  {journal} {Physical Review X}\ }\textbf {\bibinfo {volume} {13}},\ \bibinfo {pages} {041052} (\bibinfo {year} {2023})}\BibitemShut {NoStop}%
\bibitem [{\citenamefont {Ma}\ \emph {et~al.}(2023)\citenamefont {Ma} \emph {et~al.}}]{Thompson2023}%
  \BibitemOpen
  \bibfield  {author} {\bibinfo {author} {\bibfnamefont {S.}~\bibnamefont {Ma}} \emph {et~al.},\ }\bibfield  {title} {\bibinfo {title} {High-fidelity gates and mid-circuit erasure conversion in an atomic qubit},\ }\href {https://doi.org/10.1038/s41586-023-06438-1} {\bibfield  {journal} {\bibinfo  {journal} {{Nature}}\ }\textbf {\bibinfo {volume} {622}},\ \bibinfo {pages} {279–284} (\bibinfo {year} {2023})}\BibitemShut {NoStop}%
\bibitem [{\citenamefont {Yu}\ \emph {et~al.}(2025)\citenamefont {Yu} \emph {et~al.}}]{Yu2025OMG}%
  \BibitemOpen
  \bibfield  {author} {\bibinfo {author} {\bibfnamefont {Y.}~\bibnamefont {Yu}} \emph {et~al.},\ }\bibfield  {title} {\bibinfo {title} {In situ midcircuit qubit measurement and reset in a single-species trapped-ion quantum computing system},\ }\href {https://doi.org/10.1103/qfvd-93lw} {\bibfield  {journal} {\bibinfo  {journal} {Phys. Rev. Res.}\ }\textbf {\bibinfo {volume} {7}},\ \bibinfo {pages} {043355} (\bibinfo {year} {2025})}\BibitemShut {NoStop}%
\bibitem [{\citenamefont {Yang}\ \emph {et~al.}(2022)\citenamefont {Yang} \emph {et~al.}}]{Yang2022OMG}%
  \BibitemOpen
  \bibfield  {author} {\bibinfo {author} {\bibfnamefont {H.-X.}\ \bibnamefont {Yang}} \emph {et~al.},\ }\bibfield  {title} {\bibinfo {title} {Realizing coherently convertible dual-type qubits with the same ion species},\ }\href {https://doi.org/10.1038/s41567-022-01661-5} {\bibfield  {journal} {\bibinfo  {journal} {Nature Physics}\ }\textbf {\bibinfo {volume} {18}},\ \bibinfo {pages} {1058} (\bibinfo {year} {2022})}\BibitemShut {NoStop}%
\bibitem [{\citenamefont {Chen}\ \emph {et~al.}(2026)\citenamefont {Chen} \emph {et~al.}}]{Chen2025OMG}%
  \BibitemOpen
  \bibfield  {author} {\bibinfo {author} {\bibfnamefont {Z.-Y.}\ \bibnamefont {Chen}} \emph {et~al.},\ }\bibfield  {title} {\bibinfo {title} {Noninvasive mid-circuit measurement and reset on atomic qubits},\ }\href {https://doi.org/10.1103/ct8k-jgsn} {\bibfield  {journal} {\bibinfo  {journal} {Phys. Rev. A}\ }\textbf {\bibinfo {volume} {113}},\ \bibinfo {pages} {012606} (\bibinfo {year} {2026})}\BibitemShut {NoStop}%
\bibitem [{\citenamefont {Gottesman}(1997)}]{gottesman1997stabilizer}%
  \BibitemOpen
  \bibfield  {author} {\bibinfo {author} {\bibfnamefont {D.}~\bibnamefont {Gottesman}},\ }\href@noop {} {\emph {\bibinfo {title} {Stabilizer codes and quantum error correction}}}\ (\bibinfo  {publisher} {California Institute of Technology},\ \bibinfo {year} {1997})\BibitemShut {NoStop}%
\bibitem [{\citenamefont {Bravyi}\ and\ \citenamefont {Kitaev}(1998)}]{bravyi_quantum_1998}%
  \BibitemOpen
  \bibfield  {author} {\bibinfo {author} {\bibfnamefont {S.~B.}\ \bibnamefont {Bravyi}}\ and\ \bibinfo {author} {\bibfnamefont {A.~Y.}\ \bibnamefont {Kitaev}},\ }\bibfield  {title} {\bibinfo {title} {Quantum codes on a lattice with boundary},\ }\href@noop {} {\bibfield  {journal} {\bibinfo  {journal} {{P}reprint at https://doi.org/10.48550/arXiv.quant-ph/9811052}\ } (\bibinfo {year} {1998})}\BibitemShut {NoStop}%
\bibitem [{\citenamefont {Bombin}\ and\ \citenamefont {Martin-Delgado}(2007)}]{bombin_optimal_2007}%
  \BibitemOpen
  \bibfield  {author} {\bibinfo {author} {\bibfnamefont {H.}~\bibnamefont {Bombin}}\ and\ \bibinfo {author} {\bibfnamefont {M.~A.}\ \bibnamefont {Martin-Delgado}},\ }\bibfield  {title} {\bibinfo {title} {Optimal resources for topological two-dimensional stabilizer codes: {Comparative} study},\ }\href {https://link.aps.org/doi/10.1103/PhysRevA.76.012305} {\bibfield  {journal} {\bibinfo  {journal} {Physical Review A}\ }\textbf {\bibinfo {volume} {76}},\ \bibinfo {pages} {012305} (\bibinfo {year} {2007})}\BibitemShut {NoStop}%
\bibitem [{\citenamefont {{A. A. Kovalev}}\ and\ \citenamefont {{L. P. Pryadko}}(2012)}]{a_a_kovalev_improved_2012}%
  \BibitemOpen
  \bibfield  {author} {\bibinfo {author} {\bibnamefont {{A. A. Kovalev}}}\ and\ \bibinfo {author} {\bibnamefont {{L. P. Pryadko}}},\ }\bibfield  {title} {\bibinfo {title} {Improved quantum hypergraph-product {LDPC} codes},\ }in\ \href {https://doi.org/10.1109/ISIT.2012.6284206} {\emph {\bibinfo {booktitle} {2012 {IEEE} {International} {Symposium} on {Information} {Theory} {Proceedings}}}}\ (\bibinfo {year} {2012})\ pp.\ \bibinfo {pages} {348--352}\BibitemShut {NoStop}%
\bibitem [{\citenamefont {Zhang}\ \emph {et~al.}(2022)\citenamefont {Zhang} \emph {et~al.}}]{zhang_digital_2022}%
  \BibitemOpen
  \bibfield  {author} {\bibinfo {author} {\bibfnamefont {X.}~\bibnamefont {Zhang}} \emph {et~al.},\ }\bibfield  {title} {\bibinfo {title} {Digital quantum simulation of {Floquet} symmetry-protected topological phases},\ }\href {https://doi.org/10.1038/s41586-022-04854-3} {\bibfield  {journal} {\bibinfo  {journal} {Nature}\ }\textbf {\bibinfo {volume} {607}},\ \bibinfo {pages} {468} (\bibinfo {year} {2022})}\BibitemShut {NoStop}%
\bibitem [{\citenamefont {Ye}\ \emph {et~al.}(2025)\citenamefont {Ye}, \citenamefont {Wecker},\ and\ \citenamefont {Delfosse}}]{ye_beam_2025}%
  \BibitemOpen
  \bibfield  {author} {\bibinfo {author} {\bibfnamefont {M.}~\bibnamefont {Ye}}, \bibinfo {author} {\bibfnamefont {D.}~\bibnamefont {Wecker}},\ and\ \bibinfo {author} {\bibfnamefont {N.}~\bibnamefont {Delfosse}},\ }\bibfield  {title} {\bibinfo {title} {Beam search decoder for quantum {LDPC} codes},\ }\href@noop {} {\bibfield  {journal} {\bibinfo  {journal} {{P}reprint at https://doi.org/10.48550/arXiv.2512.07057}\ } (\bibinfo {year} {2025})}\BibitemShut {NoStop}%
\bibitem [{\citenamefont {Pedersen}\ \emph {et~al.}(2007)\citenamefont {Pedersen}, \citenamefont {Møller},\ and\ \citenamefont {Mølmer}}]{pedersen_fidelity_2007}%
  \BibitemOpen
  \bibfield  {author} {\bibinfo {author} {\bibfnamefont {L.~H.}\ \bibnamefont {Pedersen}}, \bibinfo {author} {\bibfnamefont {N.~M.}\ \bibnamefont {Møller}},\ and\ \bibinfo {author} {\bibfnamefont {K.}~\bibnamefont {Mølmer}},\ }\bibfield  {title} {\bibinfo {title} {Fidelity of quantum operations},\ }\href {https://doi.org/10.1016/j.physleta.2007.02.069} {\bibfield  {journal} {\bibinfo  {journal} {Physics Letters A}\ }\textbf {\bibinfo {volume} {367}},\ \bibinfo {pages} {47} (\bibinfo {year} {2007})}\BibitemShut {NoStop}%
\bibitem [{\citenamefont {Malinowski}\ \emph {et~al.}(2023)\citenamefont {Malinowski}, \citenamefont {Allcock},\ and\ \citenamefont {Ballance}}]{malinowski2023wire}%
  \BibitemOpen
  \bibfield  {author} {\bibinfo {author} {\bibfnamefont {M.}~\bibnamefont {Malinowski}}, \bibinfo {author} {\bibfnamefont {D.}~\bibnamefont {Allcock}},\ and\ \bibinfo {author} {\bibfnamefont {C.}~\bibnamefont {Ballance}},\ }\bibfield  {title} {\bibinfo {title} {How to wire a 1000-qubit trapped-ion quantum computer},\ }\href@noop {} {\bibfield  {journal} {\bibinfo  {journal} {PRX Quantum}\ }\textbf {\bibinfo {volume} {4}},\ \bibinfo {pages} {040313} (\bibinfo {year} {2023})}\BibitemShut {NoStop}%
\bibitem [{\citenamefont {L{\"o}schnauer}\ \emph {et~al.}(2025)\citenamefont {L{\"o}schnauer} \emph {et~al.}}]{loschnauer2025scalable}%
  \BibitemOpen
  \bibfield  {author} {\bibinfo {author} {\bibfnamefont {C.}~\bibnamefont {L{\"o}schnauer}} \emph {et~al.},\ }\bibfield  {title} {\bibinfo {title} {Scalable, high-fidelity all-electronic control of trapped-ion qubits},\ }\href@noop {} {\bibfield  {journal} {\bibinfo  {journal} {PRX Quantum}\ }\textbf {\bibinfo {volume} {6}},\ \bibinfo {pages} {040313} (\bibinfo {year} {2025})}\BibitemShut {NoStop}%
\bibitem [{\citenamefont {Hughes}\ \emph {et~al.}(2025)\citenamefont {Hughes} \emph {et~al.}}]{hughes2025trapped}%
  \BibitemOpen
  \bibfield  {author} {\bibinfo {author} {\bibfnamefont {A.}~\bibnamefont {Hughes}} \emph {et~al.},\ }\bibfield  {title} {\bibinfo {title} {Trapped-ion two-qubit gates with> 99.99\% fidelity without ground-state cooling},\ }\href@noop {} {\bibfield  {journal} {\bibinfo  {journal} {{P}reprint at https://arxiv.org/abs/2510.17286}\ } (\bibinfo {year} {2025})}\BibitemShut {NoStop}%
\bibitem [{\citenamefont {Tripier}\ \emph {et~al.}(2026{\natexlab{b}})\citenamefont {Tripier} \emph {et~al.}}]{walking_cat}%
  \BibitemOpen
  \bibfield  {author} {\bibinfo {author} {\bibfnamefont {F.}~\bibnamefont {Tripier}} \emph {et~al.},\ }\bibfield  {title} {\bibinfo {title} {Fault-tolerant quantum computing with trapped ions: The walking cat architecture},\ }\href@noop {} {\bibfield  {journal} {\bibinfo  {journal} {{P}reprint at https://doi.org/10.48550/arXiv.2604.19481}\ } (\bibinfo {year} {2026}{\natexlab{b}})}\BibitemShut {NoStop}%
\bibitem [{\citenamefont {Vaidman}\ \emph {et~al.}(1996)\citenamefont {Vaidman}, \citenamefont {Goldenberg},\ and\ \citenamefont {Wiesner}}]{vaidman_error_1996}%
  \BibitemOpen
  \bibfield  {author} {\bibinfo {author} {\bibfnamefont {L.}~\bibnamefont {Vaidman}}, \bibinfo {author} {\bibfnamefont {L.}~\bibnamefont {Goldenberg}},\ and\ \bibinfo {author} {\bibfnamefont {S.}~\bibnamefont {Wiesner}},\ }\bibfield  {title} {\bibinfo {title} {Error prevention scheme with four particles},\ }\href {https://link.aps.org/doi/10.1103/PhysRevA.54.R1745} {\bibfield  {journal} {\bibinfo  {journal} {Physical Review A}\ }\textbf {\bibinfo {volume} {54}},\ \bibinfo {pages} {R1745} (\bibinfo {year} {1996})}\BibitemShut {NoStop}%
\bibitem [{\citenamefont {Nielsen}\ \emph {et~al.}(2021)\citenamefont {Nielsen} \emph {et~al.}}]{Nielsen2021gatesettomography}%
  \BibitemOpen
  \bibfield  {author} {\bibinfo {author} {\bibfnamefont {E.}~\bibnamefont {Nielsen}} \emph {et~al.},\ }\bibfield  {title} {\bibinfo {title} {Gate {S}et {T}omography},\ }\href {https://doi.org/10.22331/q-2021-10-05-557} {\bibfield  {journal} {\bibinfo  {journal} {{Quantum}}\ }\textbf {\bibinfo {volume} {5}},\ \bibinfo {pages} {557} (\bibinfo {year} {2021})}\BibitemShut {NoStop}%
\bibitem [{\citenamefont {Proctor}\ \emph {et~al.}(2019)\citenamefont {Proctor} \emph {et~al.}}]{Proctor_DRB_PhysRevLett.123.030503}%
  \BibitemOpen
  \bibfield  {author} {\bibinfo {author} {\bibfnamefont {T.~J.}\ \bibnamefont {Proctor}} \emph {et~al.},\ }\bibfield  {title} {\bibinfo {title} {Direct randomized benchmarking for multiqubit devices},\ }\href {https://doi.org/10.1103/PhysRevLett.123.030503} {\bibfield  {journal} {\bibinfo  {journal} {Phys. Rev. Lett.}\ }\textbf {\bibinfo {volume} {123}},\ \bibinfo {pages} {030503} (\bibinfo {year} {2019})}\BibitemShut {NoStop}%
\bibitem [{\citenamefont {Nielsen}\ \emph {et~al.}(2020)\citenamefont {Nielsen} \emph {et~al.}}]{Nielsen_pygsti_2020}%
  \BibitemOpen
  \bibfield  {author} {\bibinfo {author} {\bibfnamefont {E.}~\bibnamefont {Nielsen}} \emph {et~al.},\ }\bibfield  {title} {\bibinfo {title} {Probing quantum processor performance with pygsti},\ }\href {https://doi.org/10.1088/2058-9565/ab8aa4} {\bibfield  {journal} {\bibinfo  {journal} {Quantum Science and Technology}\ }\textbf {\bibinfo {volume} {5}},\ \bibinfo {pages} {044002} (\bibinfo {year} {2020})}\BibitemShut {NoStop}%
\bibitem [{\citenamefont {McKay}\ \emph {et~al.}(2023)\citenamefont {McKay} \emph {et~al.}}]{mckay2023benchmarkingquantumprocessorperformance}%
  \BibitemOpen
  \bibfield  {author} {\bibinfo {author} {\bibfnamefont {D.~C.}\ \bibnamefont {McKay}} \emph {et~al.},\ }\href@noop {} {\bibinfo {title} {Benchmarking quantum processor performance at scale}} (\bibinfo {year} {2023}),\ \Eprint {https://arxiv.org/abs/2311.05933} {arXiv:2311.05933 [quant-ph]} \BibitemShut {NoStop}%
\bibitem [{\citenamefont {Bloch}(1946)}]{bloch_nuclear_1946}%
  \BibitemOpen
  \bibfield  {author} {\bibinfo {author} {\bibfnamefont {F.}~\bibnamefont {Bloch}},\ }\bibfield  {title} {\bibinfo {title} {Nuclear {Induction}},\ }\href {https://link.aps.org/doi/10.1103/PhysRev.70.460} {\bibfield  {journal} {\bibinfo  {journal} {Physical Review}\ }\textbf {\bibinfo {volume} {70}},\ \bibinfo {pages} {460} (\bibinfo {year} {1946})}\BibitemShut {NoStop}%
\bibitem [{\citenamefont {Ramsey}(1950)}]{ramsey_molecular_1950}%
  \BibitemOpen
  \bibfield  {author} {\bibinfo {author} {\bibfnamefont {N.~F.}\ \bibnamefont {Ramsey}},\ }\bibfield  {title} {\bibinfo {title} {A {Molecular} {Beam} {Resonance} {Method} with {Separated} {Oscillating} {Fields}},\ }\href {https://doi.org/10.1103/PhysRev.78.695} {\bibfield  {journal} {\bibinfo  {journal} {Physical Review}\ }\textbf {\bibinfo {volume} {78}},\ \bibinfo {pages} {695} (\bibinfo {year} {1950})}\BibitemShut {NoStop}%
\end{thebibliography}%

\pagebreak
\appendix
\crefalias{section}{appendix}

\section{Survival Curves (Toric, Concatenated Codes)}
\begin{figure}[h]
\begin{centering}
\includegraphics[width=8cm]{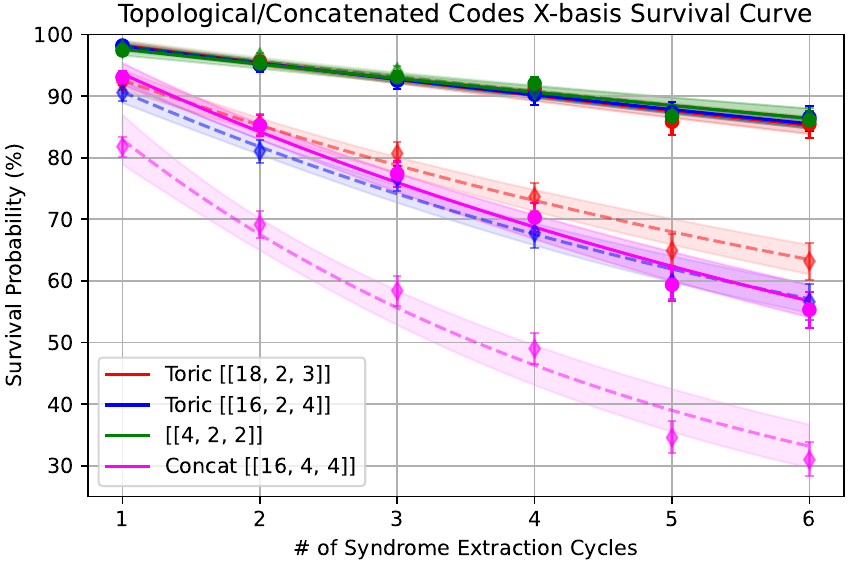}
\par\end{centering}
\caption{
Survival probabilities $1-\plog$ for $X$ eigenstate memories in the toric and concatenated codes.
}
\label{fig:Survival_Other_X}
\end{figure}

\begin{figure}[h]
\begin{centering}
\includegraphics[width=8cm]{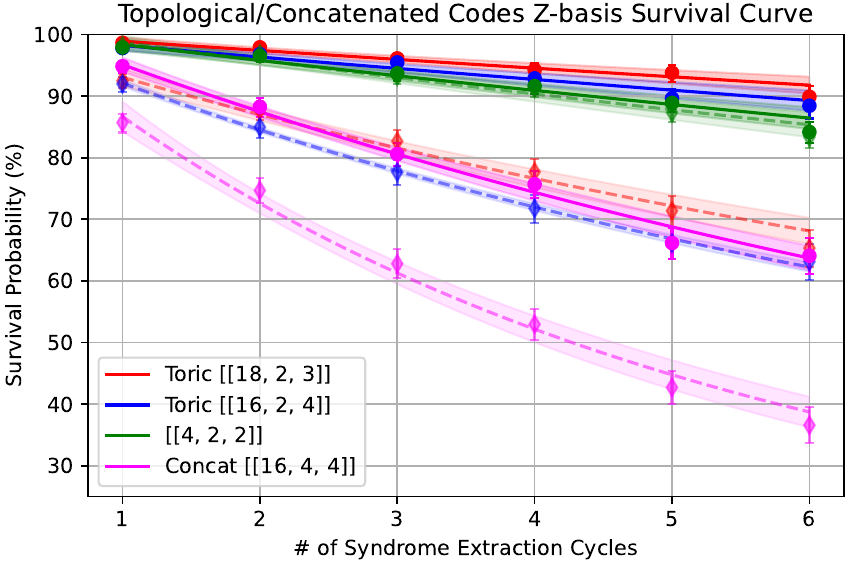}
\par\end{centering}
\caption{
Survival probabilities $1-\plog$ for $Z$ eigenstate memories in the toric and concatenated codes.
}
\label{fig:Survival_Other_Z}
\end{figure}

\Cref{fig:Survival_Other_X,fig:Survival_Other_Z} show survival curves, similar to that of \cref{fig:Survival}, for toric and concatenated codes.
The (unconcatenated) $[[4,2,2]]$ code is also shown for reference.
Note, that unlike other codes, the $[[4,2,2]]$ exhibits no improvement in logical error rate upon decoding (as is to be expected for a distance $d=2$ code).

\section{Code Definitions}
\label{app:CodePolynomials}

\subsection*{High-rate qLDPC Codes}
\begin{table}[h]
  \begin{centering}
  \begin{tabular}{|c|c|c|c|c|}
  \hline 
  Code & $\ell$ & $m$ & ${\cal A}$ & ${\cal B}$\tabularnewline
  \hline 
  \hline 
  BB$[[18,4,3]]$ & 3 & 3 & \multirow{2}{*}{$1+x$} & \multirow{2}{*}{$1+y+xy^{2}$}\tabularnewline
  \cline{1-3} \cline{2-3} \cline{3-3} 
  BB$[[24,4,4]]$ & 4 & 3 &  & \tabularnewline
  \hline 
  BB$[[30,4,5]]$ & 5 & 3 & $1+x$ & $1+y+x^{2}y^{2}$\tabularnewline
  \hline 
  GB$[[16,2,4]]$ & 1 & 8 & $1+y$ & $1+y^{5}$\tabularnewline
  \hline 
  GB$[[26,2,5]]$ & 1 & 13 & $1+y^{10}$ & $y^{9}+y^{11}$\tabularnewline
  \hline 
  \end{tabular}
  \par\end{centering}
  \caption{Table of quasi-cyclic code polynomials.}
  \label{tab:Polynomials}
\end{table}

Bivariate-Bicycle (BB)~\cite{bravyi_high-threshold_2024} and Generalized-Bicycle (GB) codes are quasi-cyclic codes built atop circulant binary matrices, $Q_\ell \in \FTwo^{\ell\times\ell}$.
This is equivalent to a square identity matrix $I_\ell$ of size $\ell$ with columns cyclicly shifted by one, whose first is $(0,1,0,0,...)$.
Then, with integers $\ell$, $m$, and matrices $x = Q_\ell \otimes I_m$, $y = I_\ell \otimes Q_m$, a BB or GB code is defined by matrix-valued polynomials $\mA(x,y)$ and $\mB(x,y)$ such that its $X$ and $Z$ stabilizer generators are taken to be first $n-k$ independent rows of:
\begin{align}
  H_X = [\mA | \mB],~&~H_Z = [\mB^T | \mA^T].
  \label{eq:HxHz}
\end{align}
The GB code differs only in that either $\ell=1$ or $m=1$.

The combined number of monomial terms that comprise $\mA$ and $\mB$ determines the weight $w_c$ of the code's stabilizer generators ({\em i.e.}, its {\em check weight}); we will refer to such a code as BB$w_c$ or GB$w_c$.
In this work, we demonstrate three BB5 instances with parameters $[[6d,4,d]]$ ($d\in\{3,4,5\}$).
We also implement two GB4 codes ($[[16,2,4]]$ and $[[26,2,5]]$).
\cref{tab:Polynomials} lists $\mA$ and $\mB$ for each.

An example Tanner graph for a BB5 circuit is shown in \cref{fig:MainFig}C in the main text.

\subsection*{Toric Codes}
\label{subsec:Toric}
Toric codes can be expressed as a specific subset of BB codes, with $\ell=m=d$ and polynomials
\begin{align}
  \mA(x) = 1+x,~&~\mB(y) = 1+y.
  \label{eq:ToricStab}
\end{align}

We implement a standard $[[18,2,3]]$ Toric code.
Additionally, we implement a rotated $[[16,2,4]]$ variant obtained by a coordinate change in $x$ and $y$~\cite{bombin_optimal_2007,a_a_kovalev_improved_2012}, with the following parity check matrices:
\begin{align}
H_X &= \left(\begin{array}{cccccccccccccccc}
1 & 1 & 0 & 0 & 1 & 1 & 0 & 0 & 0 & 0 & 0 & 0 & 0 & 0 & 0 & 0\\
0 & 0 & 0 & 0 & 0 & 1 & 1 & 0 & 0 & 1 & 1 & 0 & 0 & 0 & 0 & 0\\
0 & 0 & 0 & 0 & 0 & 0 & 0 & 0 & 1 & 1 & 0 & 0 & 1 & 1 & 0 & 0\\
0 & 1 & 1 & 0 & 0 & 0 & 0 & 0 & 0 & 0 & 0 & 0 & 0 & 1 & 1 & 0\\
0 & 0 & 1 & 1 & 0 & 0 & 1 & 1 & 0 & 0 & 0 & 0 & 0 & 0 & 0 & 0\\
0 & 0 & 0 & 0 & 1 & 0 & 0 & 1 & 1 & 0 & 0 & 1 & 0 & 0 & 0 & 0\\
0 & 0 & 0 & 0 & 0 & 0 & 0 & 0 & 0 & 0 & 1 & 1 & 0 & 0 & 1 & 1
\end{array}\right)\nonumber\\
H_Z &= \left(\begin{array}{cccccccccccccccc}
0 & 1 & 1 & 0 & 0 & 1 & 1 & 0 & 0 & 0 & 0 & 0 & 0 & 0 & 0 & 0\\
0 & 0 & 0 & 0 & 1 & 1 & 0 & 0 & 1 & 1 & 0 & 0 & 0 & 0 & 0 & 0\\
0 & 0 & 0 & 0 & 0 & 0 & 0 & 0 & 0 & 1 & 1 & 0 & 0 & 1 & 1 & 0\\
1 & 1 & 0 & 0 & 0 & 0 & 0 & 0 & 0 & 0 & 0 & 0 & 1 & 1 & 0 & 0\\
1 & 0 & 0 & 1 & 1 & 0 & 0 & 1 & 0 & 0 & 0 & 0 & 0 & 0 & 0 & 0\\
0 & 0 & 0 & 0 & 0 & 0 & 1 & 1 & 0 & 0 & 1 & 1 & 0 & 0 & 0 & 0\\
0 & 0 & 0 & 0 & 0 & 0 & 0 & 0 & 1 & 0 & 0 & 1 & 1 & 0 & 0 & 1
\end{array}\right)
\label{eq:RotatedToric}
\end{align}

\begin{figure}[h]
  \begin{centering}
  \includegraphics[width=5cm] {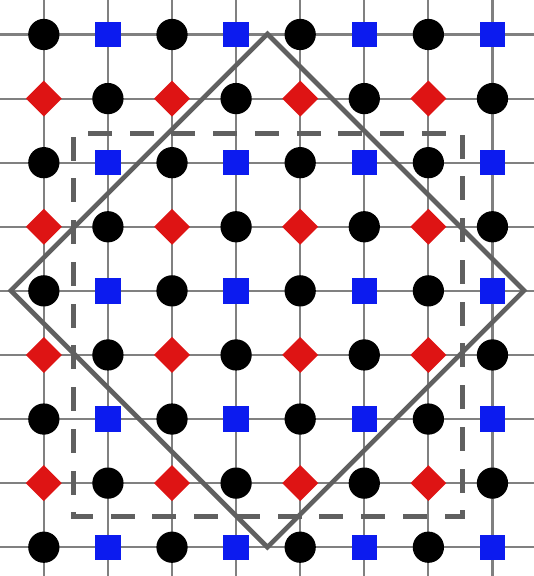}
  \par\end{centering}
  \caption{A 2D lattice on which we depict the Tanner graphs of a standard $[[18,2,3]]$ (dashed square) and rotated $[[16,2,4]]$ (rotated square) toric codes.
  Although not drawn on a toric surface, the boundaries of the solid and dashed squares are understood to wrap around along opposing edges to satisfy the toric boundary conditions.}
  \label{fig:ToricTanner}
\end{figure}

In \cref{fig:ToricTanner}, we depict a 2D lattice with
the Tanner graphs of both Toric codes we implement delimited by square boundaries. Here, toroidal boundary conditions apply between opposite edges of each square.

\subsection*{Concatenated Codes}
\label{subsec:Concat}
Code concatenation nests quantum codes by substituting physical qubits in one code with logical qubits of another code.
The result is usually a code with greater minimum distance but poorer encoding rate.

We start with the $[[4,2,2]]$ error-detection code~\cite{vaidman_error_1996}.
Its lone stabilizer generators in each basis are exceptionally simple:
\begin{align}
  S_X = X_1 X_2 X_3 X_4,~&~S_Z = Z_1 Z_2 Z_3 Z_4.
  \label{eq:ConcatStabs}
\end{align}
We then concatenate the $[[4,2,2]]$ code with itself by encoding two ``inner'' copies, which usually occupy eight physical qubits, into the logical qubits of four ``outer'' copies. This concatenation is illustrated by the Tanner graph shown in \cref{fig:MainFig}C in the main text. 

Denote by $X_{L,\kappa}^{(c)}$ and $Z_{L,\kappa}^{(c)}$ the logical Pauli operators for the $\kappa$-th logical qubit of the $c$-th ``outer'' $[[4,2,2]]$ code.
In addition to the regular stabilizer generators of each of the four ``outer'' $[[4,2,2]]$ codes, the two ``inner'' copies add the following:
\begin{align}
  S_X^{(1)} =& X_{L,1}^{(1)} X_{L,1}^{(2)} X_{L,1}^{(3)} X_{L,1}^{(4)}\nonumber\\
  S_X^{(2)} =& X_{L,2}^{(1)} X_{L,2}^{(2)} X_{L,2}^{(3)} X_{L,2}^{(4)}\nonumber\\
  S_Z^{(1)} =& Z_{L,1}^{(1)} Z_{L,1}^{(2)} Z_{L,1}^{(3)} Z_{L,1}^{(4)}\nonumber\\
  S_Z^{(2)} =& Z_{L,2}^{(1)} Z_{L,2}^{(2)} Z_{L,2}^{(3)} Z_{L,2}^{(4)}
  \label{eq:422ExtraStab}
\end{align}
The result is a $[[16,4,4]]$ concatenated code that, unlike the base $[[4,2,2]]$ code, can both detect and correct errors.

\Cref{tab:Polynomials} shows polynomials that define the parity check matrices for quasi-cyclic qLDPC codes (see \cref{eq:HxHz}).
Note that Toric codes (see \cref{subsec:Toric}) share a similar structure, but we omit them in this table since they are listed simply in \cref{eq:ToricStab}.

\section{Syndrome Circuits, State Preparation, and Measurements}
\label{app:SyndromeCircuits}

\begin{table}[h]
\begin{centering}
\begin{tabular}{|c|c|c|c|}
\hline 
Code & $n$ & $n_a$ & $n_s$ \tabularnewline
\hline 
\hline 
BB$[[18,4,3]]$ & 18 & 14 & 14 \tabularnewline
\hline 
BB$[[24,4,4]]$ & 24 & 10 & 20 \tabularnewline
\hline 
BB$[[30,4,5]]$ & 30 & 10 & 26 \tabularnewline
\hline 
GB$[[16,2,4]]$ & 16 & 14 & 14 \tabularnewline
\hline 
GB$[[26,2,5]]$ & 26 & 12 & 24 \tabularnewline
\hline 
Tor$[[18,2,3]]$ & 18 & 16 & 16 \tabularnewline
\hline 
Tor$[[16,2,4]]$ & 16 & 14 & 14 \tabularnewline
\hline 
Con$[[16,4,4]]$ & 16 & 12 & 12 \tabularnewline
\hline 
\end{tabular}
\par\end{centering}
\caption{The numbers of physical qubits in the code block ($n$), physical ancillae ($n_{a}$), and stabilizer generators measured ($n_s$) per syndrome extraction cycle used in the implementation of syndrome circuits for each code.}
\label{tab:NumAncillae}
\end{table}

Everywhere in this work, we use bare ancillae syndrome extraction circuits.
Due to the way our system is engineered, gates are applied strictly sequentially.
Gates follow a ``check first'' order -- that is, all gates for one check are applied before gates for a subsequent check -- as illustrated in \cref{fig:MainFig}B.

We interleave $X$ and $Z$ stabilizer generators -- {\em i.e.}, $X$ stabilizer generators are always followed by $Z$ stabilizer generators, and vice versa.
Denote by $\Hxred$ and $\Hzred$ the matrices consisting of the first $(n-k)/2$ independent rows of $H_X$ and $H_Z$, respectively (see \cref{eq:HxHz,eq:ToricStab,eq:RotatedToric}).
And denote by $\hxred[i]$ and $\hzred[i]$ the $i$-th row of $\Hxred$ and $\Hzred$, respectively.
For each $i$, we execute a controlled-Pauli gate controlled by an ancilla, targeting qubits at indices $j$ where $\hxred[i][j]$ or $\hzred[i][j]$ is nonzero; we do so in ascending order of $j$.
Then, our syndrome circuits for one syndrome extraction cycle of qLDPC and Toric codes measure checks in the following order: $\hxred[1], \hzred[1], \hxred[2], \hzred[2],...,\hxred[(n-k)/2],\hzred[(n-k)/2]$.
We checked numerically that such a construction results in fault-tolerant syndrome circuits.

In the case of the $[[4,2,2]]$ code, we simply implement $S_X$ first before $S_Z$ (see \cref{eq:ConcatStabs}).
In its concatenated variant, we implement $S_X$ and $S_Z$ for each $[[4,2,2]]$ block first (interleaving $X$ and $Z$ as before), before proceeding with the weight-8 stabilizer generators in \cref{eq:422ExtraStab}.
Notably, these weight-8 checks result in a non-fault-tolerant circuit for the concatenated $[[4,2,2]]$ code.

In \cref{tab:NumAncillae}, we list the numbers of physical qubits in the code block ($n$), physical ancillae ($n_a$), and stabilizer generators measured in each syndrome extraction cycle ($n_s = n-k$).
Whereas $n$, $k$, and $n_s$ are attributes of the code, we select $n_a$ under the constraint of only having a total of 40 physical ions.
In GB$[[16,2,4]]$ and BB$[[18,4,3]]$, we select $n_a=n_s$, so the number of MCM rounds is equal to $\nSEC-1$ (with stabilizer generators in the last syndrome extraction cycle measured via one final destructive measurement that includes all $n_a$ ancillae).
In GB$[[26,2,5]]$ and BB$[[24,4,4]]$, we select $n_a=n_s/2$, so there are $2\nSEC-1$ MCM rounds.

Finally, in BB$[[30,4,5]]$, we use only $n_a=10$ ancillae, fully utilizing our 40-ion chain.
However, because $n_s$ and $n_a$ are not congruent, we make use of the ancilla pipelining technique described in the main text to minimize the number of MCM rounds required. For example, for $\nSEC=3$, we require 7 MCM rounds (plus a set of final destructive measurements) to measure all $\nSEC\times n_s=78$ required stabilizer generators using our $n_a=10$ ancillae.
In this example, we save one MCM round by fully utilizing the third and fourth rounds to read out ancillae corresponding to consecutive syndrome extraction cycles.

\section{Quantum Characterization, Verification, and Validation (QCVV)}
\label{app:QCVV}
We intersperse QEC circuit experiments with characterization circuits designed to assess the level and type of errors on the qubits.

One- and two-qubit gates in the system are periodically calibrated using long-standing techniques which, with high confidence, eliminate systematic errors in the control pulses. 
We therefore assume predominantly stochastic noise on these gates, so we use direct randomized benchmarking (DRB) to determine the overall error rate as explained below.

The behaviour of MCM operations is less established, given the novelty of our OMG-based approach.
In particular, we want to quantify the effective idle channel afflicting qubits of the code block during the MCM protocol.
We do so with two standard characterization protocols: gate set tomography (GST) \cite{Nielsen2021gatesettomography} and direct randomized benchmarking (DRB) \cite{Proctor_DRB_PhysRevLett.123.030503}.
In each case, we apply the protocol to a single target qubit, on which gates from the set $G\in \{R_X(\pi/2), R_Y(\pi/2), I\}$ are applied (here $I$ denotes ``idle'', with no operation being applied to the target qubit).
During each $I$ operation, we perform one round of MCM during which a fixed set of \emph{different} qubits is read out.
We reiterate, though, that the entire chain is shelved to $\Dqb$ during the MCM procedure, as it is during computations.
That way, errors accumulated throughout this $I$ operation correspond to idle noise accrued by the target qubit as a result of an MCM round.

GST by design estimates errors on each operation in $G$ separately, and so obtaining idle noise using GST requires no special modification to the standard protocol.
The (single-qubit) GST noise model is as follows:
four parameters specify coherent and stochastic errors on the laser amplitude and phase used to implement the $R_X(\pi/2)$ and $R_Y(\pi/2)$ gates, and
idle noise is parameterized using 1) the rates of coherent X, Y, and Z rotations, and 2) the rates of X-, Y-, and Z-axis Pauli stochastic errors.
The noise model therefore has a total $4 + 6 = 10$ parameters.

We select fiducial and germ circuits using the \texttt{pygsti} package~\cite{Nielsen_pygsti_2020}, yielding four germs (the repeated portion of a GST circuit) and 14 fiducial pairs (the non-repeated beginning and ending portions of a GST circuit).
Germ depths of $1$, $2$, $4$, $8$, $16$, and $32$ result in a total of $81$ GST circuits.
We initially ran GST with variations on how the MCM was performed in order to tune the MCM operation (varying delay times and laser pulse parameters).
It consistently achieved a good fit to the data ($n_\sigma < 3$, a metric used in \cite{Nielsen2021gatesettomography}), indicating its error estimates could be trusted to provide useful information during this tune-up stage.
Throughout all the runs, estimated infidelities of the single-qubit gates ($R_X(\pi/2), R_Y(\pi/2)$) was $\approx 3\times 10^{-4}$.
After tuning of MCM was complete, the estimated entanglement infidelity of idle noise (due to the MCM's effect on the target qubit) was between $0$ and $8 \times 10^{-4}$.
The idle noise was predominantly Z-axis stochastic error, which is expected given the underlying physics of the system.

As a further check of the overall effect an MCM round has on data qubits, we perform DRB.
DRB circuits are constructed by randomly sampling gates~\cite{Proctor_DRB_PhysRevLett.123.030503}.
We sample gates from the set $G$, with probability $q(g)$ for $g \in G$.
We let $q(R_X(\pi/2)) = q(R_Y(\pi/2)) = \frac{1}{2}(1-q(I))$, and we generate circuits with depths of 1, 50, 300, and 1000, each with $q(I) \in \{0.25, 0.75\}$.
For each $q(I)$, the survival curve with respect to circuit depth is fitted, and the two resulting curves are used to infer different error rates for $I$ and \{ $R_X$ or $R_Y$ \}.
From DRB, we infer an idle noise of $8.7 \times 10^{-4} \pm 7.2 \times 10^{-4}$ during an MCM round.
This is consistent with the GST result.

\begin{figure}
\begin{centering}
  \includegraphics[width=0.95\linewidth]{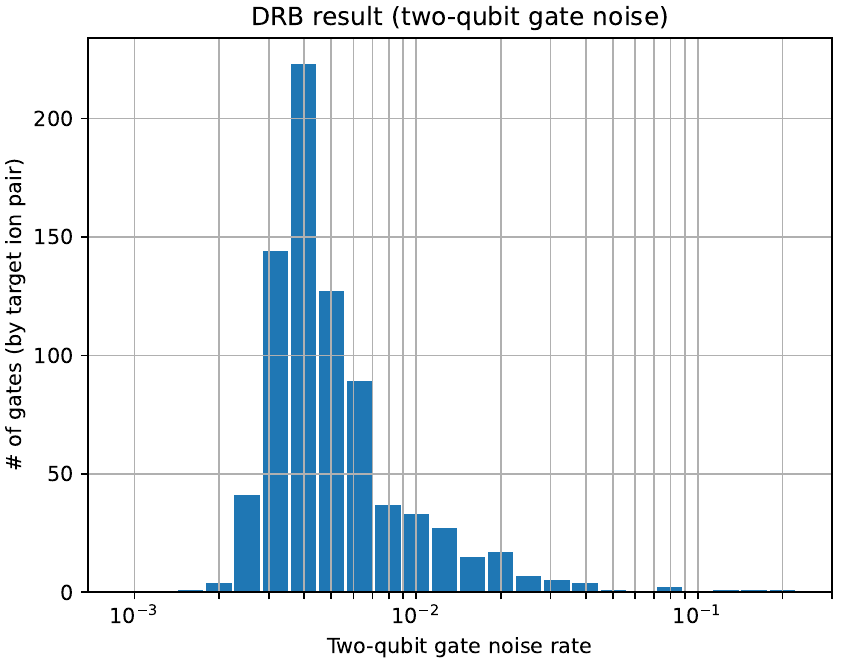}
\end{centering}
\caption{Distribution of two-qubit (MS) gate noise, as measured by DRB.}
\label{fig:2QDRB}
\end{figure}

We similarly use DRB to obtain error rates for two-qubit Mølmer-Sorensen (MS) gates -- the native two-qubit gate on our device.
Each DRB experiment (targeting MS gates on one pair qubits) consists of 4 random DRB circuits at depths 1, 5, 22, and 100.
Within each layer of the DRB circuit, we select either the MS gate with probability $q(MS)$, or a layer of single-qubit gates otherwise.
Each single-qubit gate layer consists of gates chosen uniformly randomly from $\{R_X(\pi/2), R_Y(\pi/2)\}$.
We let $q(MS) \in \{0.25, 0.75\}$.
We perform the DRB experiment for all possible pairs of target qubits on a 40-ion chain, which we refer to as a {\em fully-spanning two-qubit DRB}.
\Cref{fig:2QDRB} shows the distribution of noise rates we measured.
We use this distribution to select qubit-to-ion mappings such that high-performing MS gates are preferentially used while low-performing MS gates are avoided (see \cref{app:Qubit Mapping}).

\begin{figure}
\begin{centering}
  \includegraphics[width=0.95\linewidth]{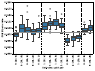}
\end{centering}
\caption{Simultaneous two-qubit DRB results.  The estimated MS gate error rate, aggregated over 36 runs, is shown as a function of qubit pair.  In each run, simultaneous DRB is performed on the sets of 5, 4, 3, and then 2 disjoint pairs separated by the vertical dashed lines.  An average error rate above the mean in Fig.~\ref{fig:2QDRB}, indicated by the horizontal line, indicates the presence of context dependency.}
\label{fig:LayerDRB}
\end{figure}

The two-qubit DRB described above measures the noise rate of one MS gate ({\em i.e.}, targeting one pair of qubits) at a time, in isolation.
As a consequence, the results shown in \cref{fig:2QDRB} leave out other context-dependent sources of errors that might degrade MS gates during regular circuit execution.
Therefore, we supplemented the fully spanning two-qubit DRB above with {\em simultaneous two-qubit DRB}~\cite{mckay2023benchmarkingquantumprocessorperformance}.
In simultaneous two-qubit DRB, circuits are designed as described above, except instead of targeting just one pair of qubits at a time, each circuit applies gates to a small set of disjoint qubit pairs.
We select representative sets on 5 qubit pairs (10 qubits) and perform two-qubit DRB on 2, 3, 4, or all 5 of the pairs simultaneously.
(We note that even though the DRB circuits are constructed with layers of gates comprised of multiple non-overlapping two-qubit gates, in practice those two-qubit gates are still executed sequentially due to hardware constraints).
\Cref{fig:LayerDRB} shows the two-qubit gate noise rates (indexed by qubit pair) obtained from simultaneous two-qubit DRB.

We find that simultaneous two-qubit DRB yields higher noise rates than the fully-spanning DRB (see \cref{fig:2QDRB}) that targeted only one qubit pair at a time. We attribute this increase to context-dependent effects that are generally not reflected in the fully-spanning DRB.
The noise rates obtained through simultaneous DRB reflect two-qubit gate performance under more realistic conditions.
Since it would take a prohibitively long time to run simultaneous DRB on all possible sets of target qubit pairs, we instead scale noise rates from fully-spanning DRB (\cref{fig:2QDRB}, omitting pairs we do not use in the experiment) so that its mean matches the average noise rate from small sets like that shown in \cref{fig:LayerDRB}.
We use that scaled noise rate to inform the two-qubit gate noise we use to instantiate our decoder's detector error model.
Furthermore, by targeting only small sets of qubit pairs, our simultaneous DRB experiments are fast enough to run periodically between QEC memory experiments, enabling them to serve as point checks to catch anomalous system behaviour.

\section{Approximate Pauli Noise Model}
\label{app:Noise Model}

\begin{table}[h]
  \begin{centering}
  \begin{tabular}{|c|c|c|}
  \hline 
  Noise & Type & Strength\tabularnewline
  \hline 
  \hline 
  \multirow{2}{*}{Idle ($\Sqb$)} & $X$ error & $10^{-5}$\tabularnewline
  \cline{2-3} \cline{3-3} 
  & $Z$ error & $4\times10^{-5}$\tabularnewline
  \hline 
  \multirow{2}{*}{Idle (per MCM round)} & $X$ error & $10^{-4}$\tabularnewline
  \cline{2-3} \cline{3-3} 
  & $Z$ error & $4\times10^{-4}$\tabularnewline
  \hline 
  Two-qubit gate & Depolarizing & {*}\tabularnewline
  \hline 
  One-qubit gate & Depolarizing & $3.2\times10^{-4}$\tabularnewline
  \hline 
  Measurement & Bit-flip & $9\times10^{-3}$\tabularnewline
  \hline 
  Leakage (per MCM round) & - & $1.7\times10^{-3}$\tabularnewline
  \hline 
  \end{tabular}
  \par\end{centering}
  \caption{Table of stochastic Pauli noise.
  Here, asterisk ({*}) indicates a variable value (see \cref{fig:2QDRB})}
  \label{tab:PauliNoise}
\end{table}

\Cref{tab:PauliNoise} shows the stochastic Pauli noise model used for simulations and for instantiating our decoder.
We obtain these values first with a coarse estimate via QCVV (\cref{app:QCVV}), and then with fine tuning by repeatedly decoding under slightly different values for each noise.
During fine tuning, we use the sum of all logical error rates across all memory experiments as our cost function.

Idle noise in $\Sqb$ is estimated per gate time (for simplicity, we do {\em not} assign different durations to two- and one-qubit operations).
We ascribe different $X$ vs $Z$ error rates, to reflect known behaviour of our physical qubits in both $\Sqb$ and $\Dqb$ manifolds, with $Z$ errors (phase-flips) being far more prevalent compared to $X$ errors (bit-flips).
Idle noise during mid-circuit measurement (MCM) encompasses the full shelve-measure-deshelve-reset sequence described in \cref{sec:methods}, and we model it as a single layer of idle operations during which every qubit that is not measured ({\em i.e.}, not an ancilla) is afflicted by depolarizing noise with the stated rate.

We assume leakage occurs only during MCM.
For decoding of experimental data, the decoder is instantiated ignoring leakage error since we post-select away instances where leakage is detected.
For simulations, on the other hand, where erasure conversion is assumed (see \cref{fig:SimGB4,fig:SimGB4}), we insert fully-depolarized states with the stated probability on each qubit that is not measured, immediately following each round of MCM.

\section{Leakage, Post-selection, and Simulation}
\label{app:Postselect}
To avoid deleterious effects of subspace leakage, we post-select and reject shots in which leakage was detected (see the Mid-Circuit Measurement section of the main text).
The dominant source of leakage on our system is technical imperfection in the mid-circuit measurement (MCM) protocol.
Therefore, increasing the number $\nSEC$ of syndrome extraction cycles, which necessitates more MCM rounds, correlates strongly with increased leakage.

\begin{figure}[H]
\begin{centering}
  \includegraphics[width=0.95\linewidth]{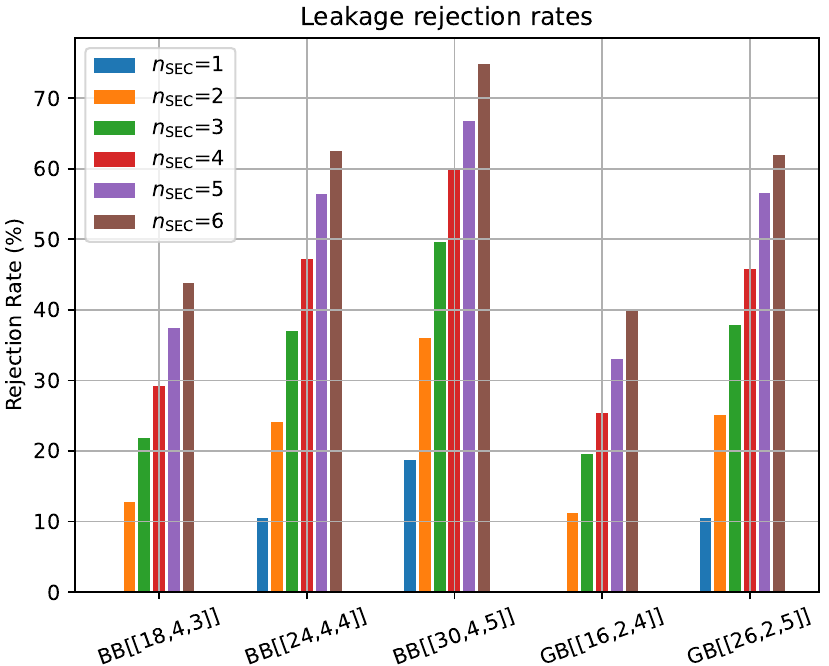}
\end{centering}
\caption{Leakage detection and rejection rates for various codes and number of syndrome extraction cycles $\nSEC$.}
\label{fig:LeakagePS}
\end{figure}

\Cref{fig:LeakagePS} shows the resulting rejection rates for our high-rate qLDPC code implementations.
The BB$[[30,4,5]]$ code, which was most demanding in number of MCM rounds, exhibited the highest rejection rates with $\nSEC=1$ and $\nSEC=6$ syndrome extraction cycle implementations being rejected at rates of 18.8\% and 74.8\%, respectively.

\begin{figure}[h]
\begin{centering}
\includegraphics[width=7.5cm]{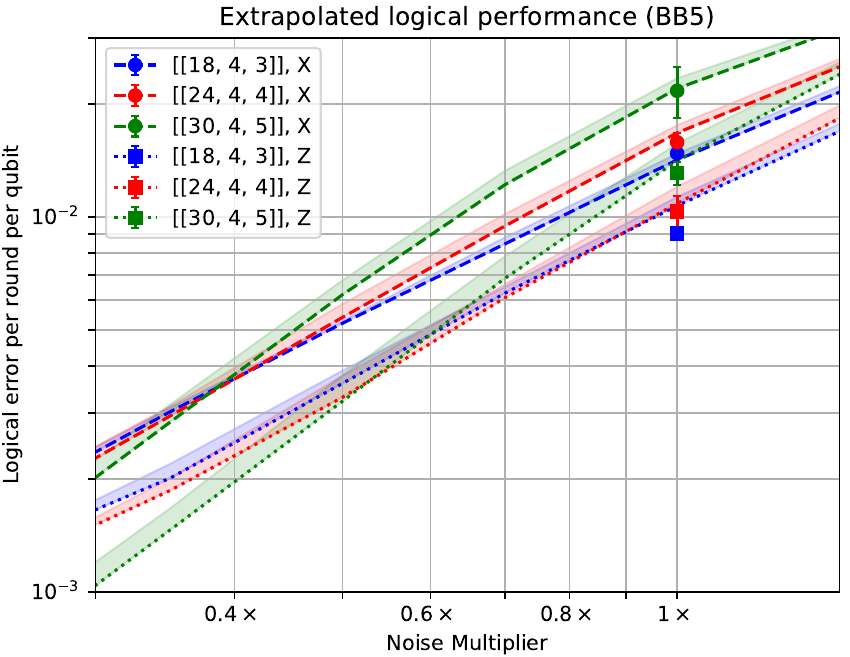}
\par\end{centering}
\caption{Simulated BB5 memory performance with increased/decreased physical noise rates.
Dots along $1\times$ noise multiplier are experimental points for reference.
Shaded bands show increase in logical error rate if leaked qubits are re-initialized with the maximally mixed state instead of rejected via post-selection.
In the worst case, erasure conversion yields relative increase of $11\%$ (dotted, green).}
\label{fig:SimBB5}
\end{figure}

\begin{figure}[h]
\begin{centering}
\includegraphics[width=7.5cm]{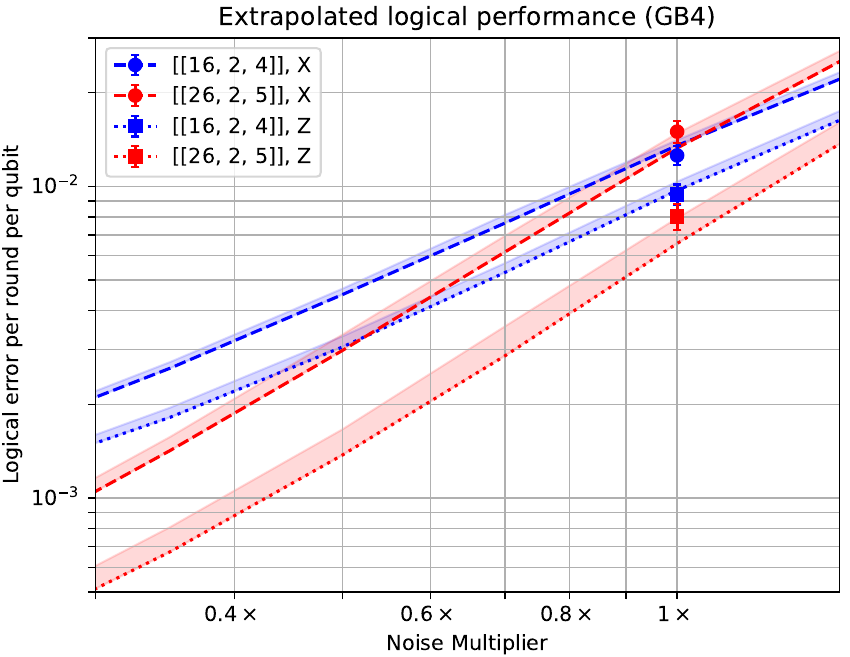}
\par\end{centering}
\caption{Simulated GB4 memory performance with increased/decreased physical noise rates.
Dots along $1\times$ noise multiplier are experimental points for reference.
Shaded bands show increase in logical error rate if leaked qubits are re-initialized with the maximally mixed state instead of rejected via post-selection.
In the worst case, erasure conversion yields relative increase of $20\%$ (dotted, red).}
\label{fig:SimGB4}
\end{figure}

Though we do not do so in this work, it is possible to convert detected leakage into erasures, avoiding post-selection entirely.
We evaluate the impact such an erasure conversion can have on logical performance through simulation.
\Cref{fig:SimBB5,fig:SimGB4} show simulation results using the same stochastic Pauli noise used to instantiate our decoder (see \cref{app:Noise Model}).
In these simulations, we scale every noise source in the memory circuit by a fixed amount, shown as ``multipliers'' on the horizontal axes.
Points at the $1\times$ multiple column show experimental data, for comparison.

Shaded bands above each curve in \cref{fig:SimBB5,fig:SimGB4} show corresponding increases in logical error rates resulting from using erasure conversion instead of post-selection.
To obtain those shaded bands, we simulate the case where each detected leakage is followed by reset and re-initialization of the afflicted qubit(s) in the maximally mixed state.

In the worst case, the relative increase in logical error rate from that added erasure amounts to $11\%$ for BB5 codes, or equivalently a $10\%$ decrease in logical qubit survival time $T_Z$.
For GB4 code, it amounts to a $20\%$ relative increase in logical error rate, or a $17\%$ decrease in logical qubit survival time $T_Z$, at worst.

\section{Qubit Mapping}
\label{app:Qubit Mapping}
Due to technical imperfections ({\em e.g.}, in calibration or gate design), not all ion pairs experience the same two-qubit gate fidelity.
Therefore, for a given circuit with gates performed between a fixed set of qubit pairs, the mean infidelity of the gates in that circuit depends on how those qubit pairs are mapped to ion pairs.
Because our Raman architecture enables us to apply two-qubit gates between any pair of ions, we can, in principle, specify an arbitrary qubit-to-ion mapping.
We note that such flexibility is not shared by hardware with fixed qubits, wherein the set of permissible two-qubit gates is sparse.

In this work, we choose a slightly constrained approach for the sake of technical simplicity. For each circuit code, we select disjoint sets of ions to serve as readout ancillae and non-ancilla data qubits. These sets are both contiguous, so that all ancillae are grouped together and all data qubits are grouped together. We emphasize that this restriction is a choice and not a fundamental limitation of our system. We then vary the mapping between ancillae and the first set of ions and the mapping between data qubits and the second set of ions.

To guide this remapping, we extract our figure of merit from the fully-spanning direct randomized benchmarking (DRB) data described in \cref{app:QCVV}. For a given code and mapping, we calculate the average gate infidelity of all two-qubit gates in the code's syndrome circuit, weighted by the number of times each gate is used. Starting from several random initial mappings, we iteratively select the qubit pair that contributes the most infidelity and try to reduce or eliminate its impact by varying the ancilla and data qubit mappings.

\begin{figure}[H]
\begin{centering}
\includegraphics[width=\linewidth]{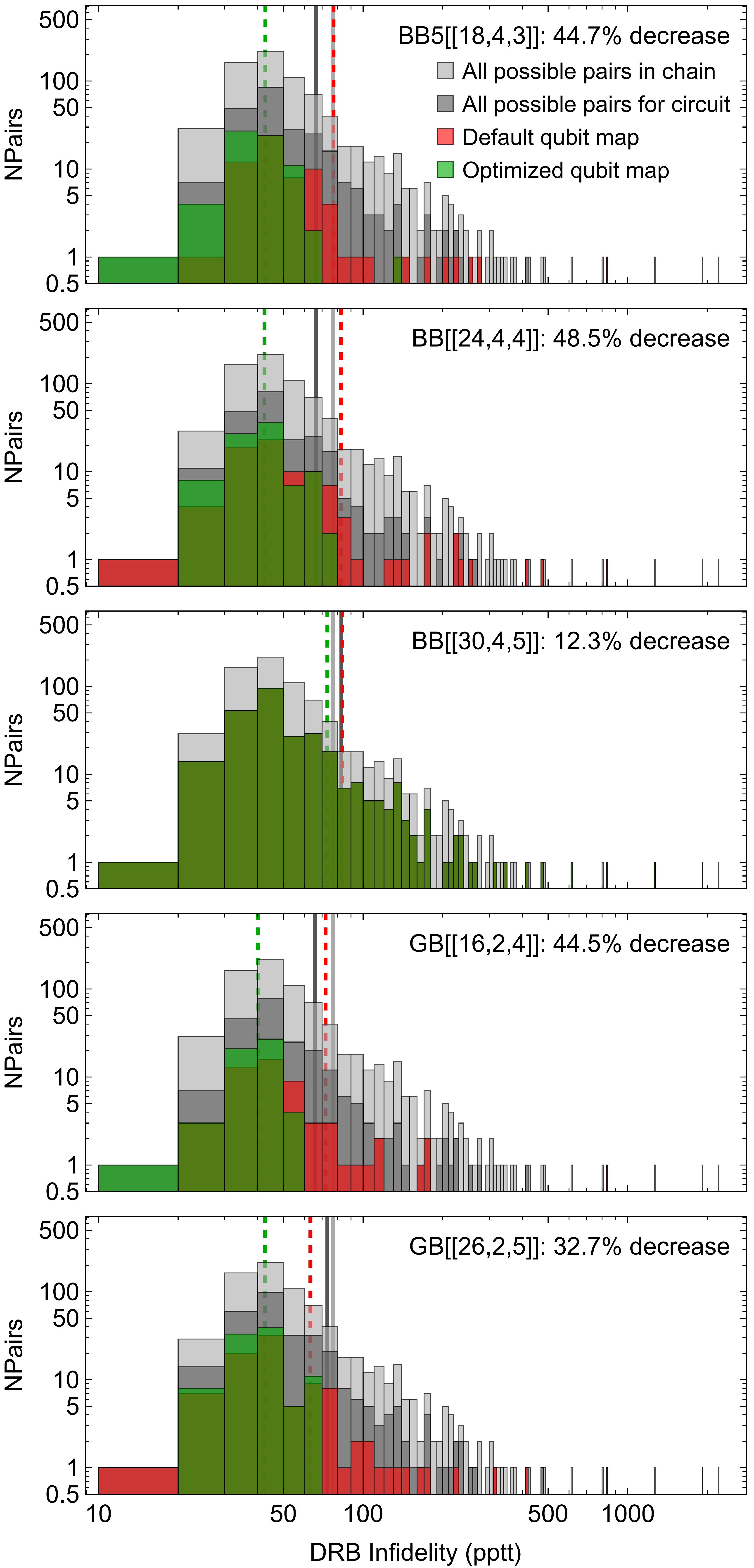}
\par\end{centering}
\caption{Distribution of DRB fidelity of two-qubit gates utilized in each circuit. The dashed lines are placed at the mean infidelity (unweighted for the grey histograms, weighted by the number of times each ion pair is used in the syndrome circuits for the red and green histograms). The infidelity reduction percentages are calculated for the weighted mean of the optimized map (green) relative to that of the default map (red).}
\label{fig:QubitIonMap}
\end{figure}

The results of this remapping are illustrated in \cref{fig:QubitIonMap}, which shows a series of histograms of two-qubit gate infidelities for each high-rate qLDPC code in our experiment. The lighter grey histograms show all 780 ion pairs that are possible in a 40-ion chain, and the darker grey histograms show only the subset of pairs, which consist of one ion from the set of ancilla ions and one from the set of data qubit ions, that could possibly be used for each code. As each code might require different numbers of ancillae and data qubits, the darker grey histograms may vary slightly. The red histograms show the pairs corresponding to an arbitrary default map, whereas the green histograms show the pairs corresponding to an optimized map. We do not claim that these maps are the globally optimal choices, but they do reduce the average infidelity, relative to the default maps, by approximately a third to a half for most circuits.

We realize this reduction in mean infidelity because this algorithm has the effect of avoiding ion pairs in the higher-error end of the fidelity distribution, instead concentrating the ion pairs used by the circuits in the lower-error end of the distribution. In other words, we select a bespoke mapping for each code that replaces low-quality entangling gates with higher-quality gates, with no change to the system hardware, gate calibrations, or circuit structure.

The BB$[[30,4,5]]$ code is a notable exception. As shown in \cref{tab:NumAncillae}, this code is the only one for which the number of MCM bits to be read out per round is not an integer multiple of the number of physical readout ancillae used. As we describe in \cref{app:SyndromeCircuits}, we pipeline ancilla usage to reduce the number of MCM rounds required, which means that a single MCM round may be used to measure stabilizer generators from consecutive syndrome extraction cycles. As a result, for this particular code, the same stabilizer generator can be read out using different ancillae in different rounds, and so this code makes use of every possible ion pair at some point during its multiple syndrome extraction cycles. This gives us less flexibility to remap this code, and so the remapping gains (12\% infidelity reduction compared to a default map) are more modest; we can choose to use less-performant ion pairs fewer times during the syndrome circuit, but, unlike with the other codes, we cannot avoid them altogether. We note, however, that we could choose not to pipeline ancilla usage, in which case the improvement due to remapping would be comparable to those for the other codes, but that would require a greater number of MCM rounds for some numbers of syndrome extraction cycles.

\section{Syndrome Circuit Duration}
\label{app:Syndrome Circuit Timing}

\begin{figure*}
\begin{centering}
\begin{tabular}{cc}
\tabularnewline
\includegraphics[width=8.5cm]{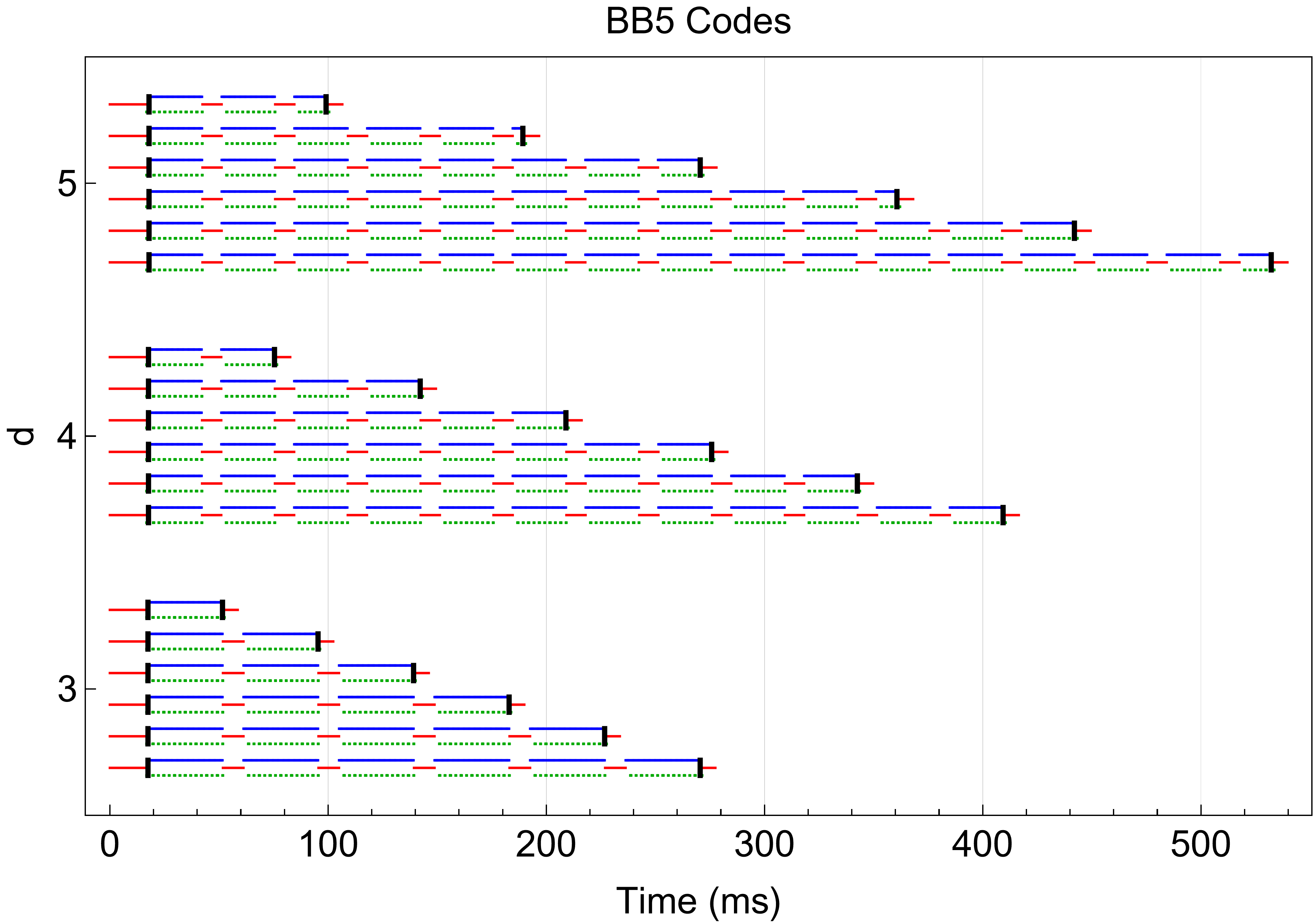}
\hspace{0.2cm}
\includegraphics[width=8.5cm]{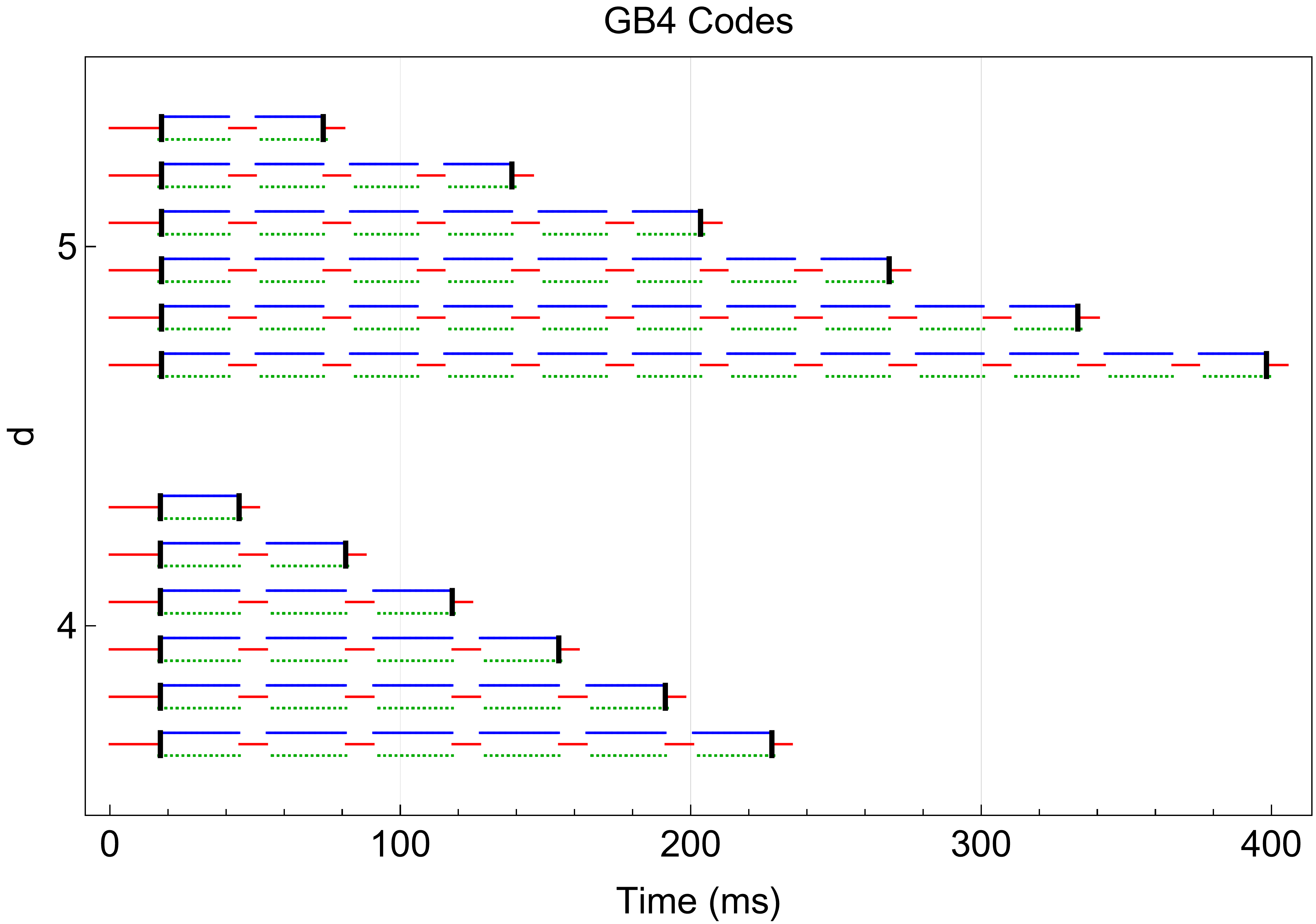}
\vspace{0.2cm}\tabularnewline
\includegraphics[width=8.5cm]{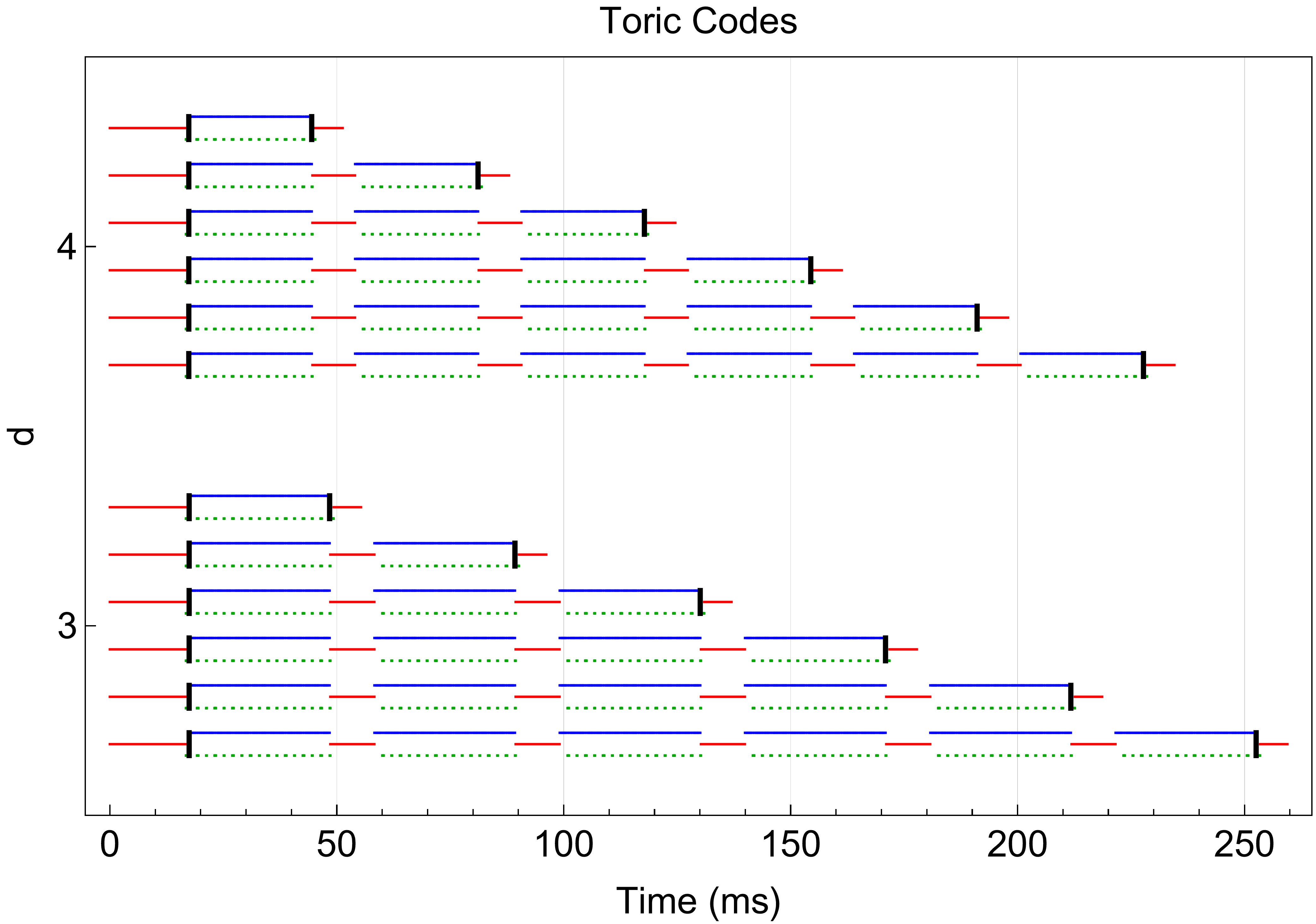}
\hspace{0.2cm}
\includegraphics[width=8.5cm]{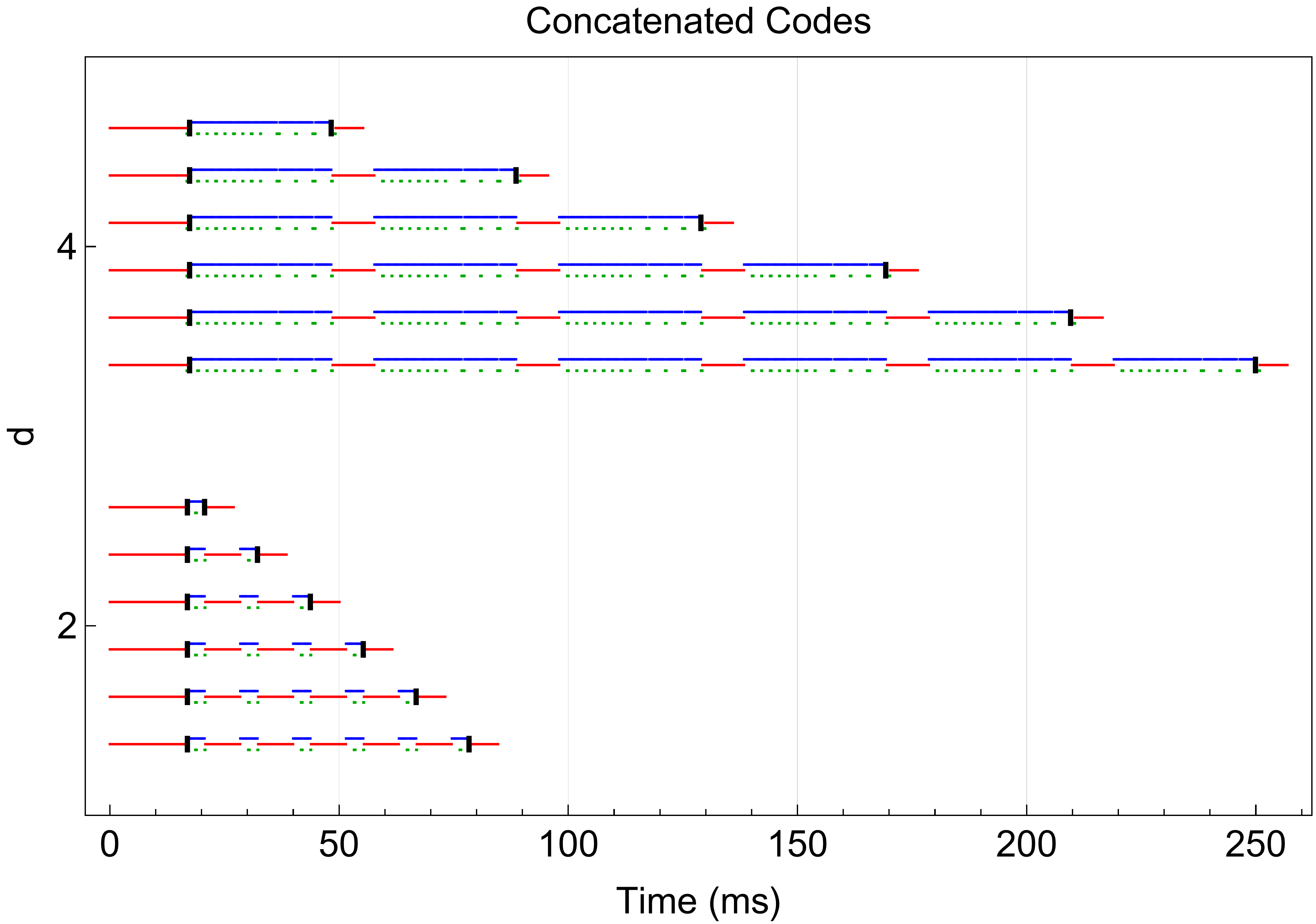}\tabularnewline
\end{tabular}
\par\end{centering}
\caption{Green/blue lines indicate one-/two-qubit Raman gates, and red lines indicate periods of state preparation, MCM, or readout. All Raman gates are plotted individually, but the lines overlap because of finite line thickness. Black lines indicate the circuits' start and end times. Circuits are grouped by value of $d$ (code distance), with the number of syndrome extraction cycles $\nSEC$ running from 1-6 from top to bottom within each group.}
\label{fig:CircuitPulses}
\end{figure*}

In this section, we show a detailed breakdown of execution time for the syndrome circuits for all codes.
We depict the full circuits graphically in \cref{fig:CircuitPulses}.
For each code instance, we show six horizontal traces corresponding to implementations over $\nSEC=1,...,6$ syndrome extraction cycles.
Vertical black bars delimit the start and stop of the memory experiment. The leading/trailing red traces correspond to the initial state preparation of physical qubits and their final state measurement, which we do not count when computing the ``lifetime'' of our logical qubits (see main text). We also neglect the single-qubit gates required to prepare the logical qubits into the various $X_L$ or $Z_L$ eigenstates.

The durations of the syndrome circuit segments (green/blue) in between MCM rounds (red) are generally proportional to $w n_a$, with $w$ being the stabilizer generator weight of the code ($w=4$ and $w=5$ for the GB and BB codes we implement, respectively) and $n_a$ denoting the number of physical ancillae in the implementation.

The relationship between the number of syndrome extraction cycles $\nSEC$ and the number of measurement segments (red; includes MCM rounds and the final destructive measurement) is discussed in \cref{app:SyndromeCircuits}. Most codes require either $\nSEC-1$ or $2\nSEC-1$ MCM rounds, depending on whether we read out all ($n_a=n_s$) or half ($n_a=n_s/2$) of our stabilizer generators in each MCM round. However, due to ion availability, the BB$[[30,4,5]]$ code uses just $n_a=10$ ancillae to measure $n_s=26$ stabilizer generators per syndrome extraction cycle. Because $n_a$ and $n_s$ are not commensurate, our pipelining of ancillae (see \cref{app:SyndromeCircuits}) implies the exact number of red measurement segments (given by $\lceil \nSEC(n-k)/{n_a} \rceil$) is not a constant multiple of $\nSEC$. Another consequence is that the final syndrome circuit segment for this code is often shorter than the other segments, and its length varies from one value of $\nSEC$ to another.

\begin{figure}[H]
\begin{centering}
\includegraphics[width=\linewidth]{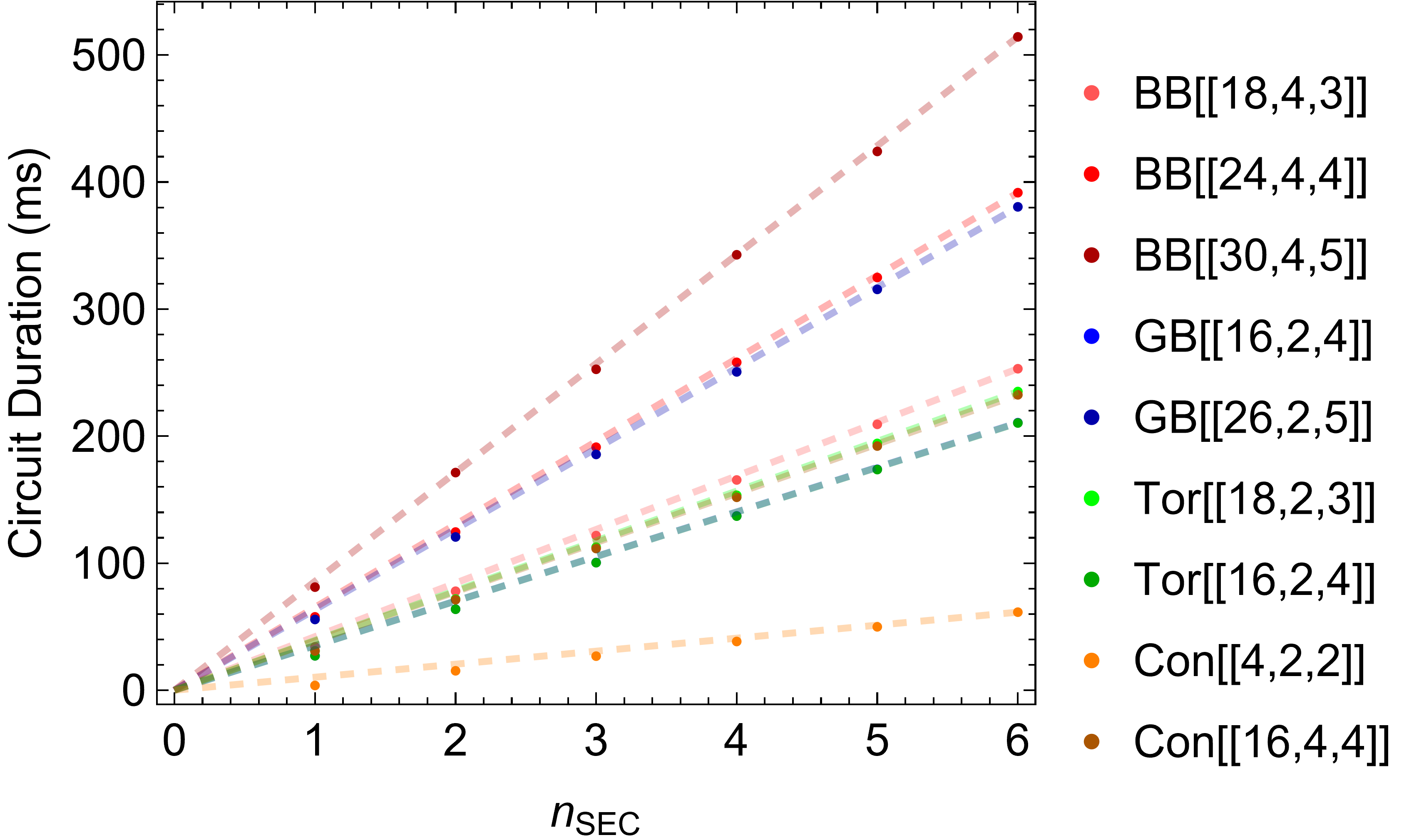}
\par\end{centering}
\caption{Durations of all qLDPC codes implemented. Dashed lines are $\nSEC\times T_\mathrm{cycle}$, included as guides to the eye, where $T_\mathrm{cycle}$ are the cycle times listed in \cref{tab:CycleTimes}.}
\label{fig:CircuitDurations}
\end{figure}

\Cref{fig:CircuitDurations} shows the same memory experiment durations as \cref{fig:CircuitPulses}, but as a function of $\nSEC$. We define the cycle time $T_\mathrm{cycle}$ for each code to be the duration of the longest memory experiment for that code ({\em i.e.}, that of $\nSEC=6$), divided by six. We list these values in \cref{tab:CycleTimes}.
We show linear interpolations based on $T_\mathrm{cycle}$ in \cref{fig:CircuitDurations} as guides to the eye, primarily to emphasize the fact that the duration of each memory experiment is well approximated by a linear function in $\nSEC$.
To obtain a survival time for logical qubits with each code, we multiply the value $\tau$, which we extract from the exponential fit to that code's observed logical error rate (see \cref{eq:Fit}), by that code's cycle time $T_\mathrm{cycle}$.

\begin{table}
  \begin{centering}
  \begin{tabular}{|c|c|}
  \hline 
  Code & $T_\mathrm{cycle}$ (ms)\tabularnewline
  \hline 
  \hline 
  BB$[[18,4,3]]$ & 42.18 \tabularnewline
  \hline 
  BB$[[24,4,4]]$ & 65.27 \tabularnewline
  \hline 
  BB$[[30,4,5]]$ & 85.71 \tabularnewline
  \hline 
  GB$[[16,2,4]]$ & 35.09 \tabularnewline
  \hline 
  GB$[[26,2,5]]$ & 63.41 \tabularnewline
  \hline 
  Tor$[[18,2,3]]$ & 39.18 \tabularnewline
  \hline 
  Tor$[[16,2,4]]$ & 35.06 \tabularnewline
  \hline 
  Con$[[16,4,4]]$ & 38.75 \tabularnewline
  \hline 
  \end{tabular}
  \par\end{centering}
  \caption{Cycle time $T_\mathrm{cycle}$ (\emph{i.e.}, duration per syndrome extraction cycle) for each code, in milliseconds.}
  \label{tab:CycleTimes}
\end{table}

\begin{figure}[H]
\begin{centering}
\includegraphics[width=\linewidth]{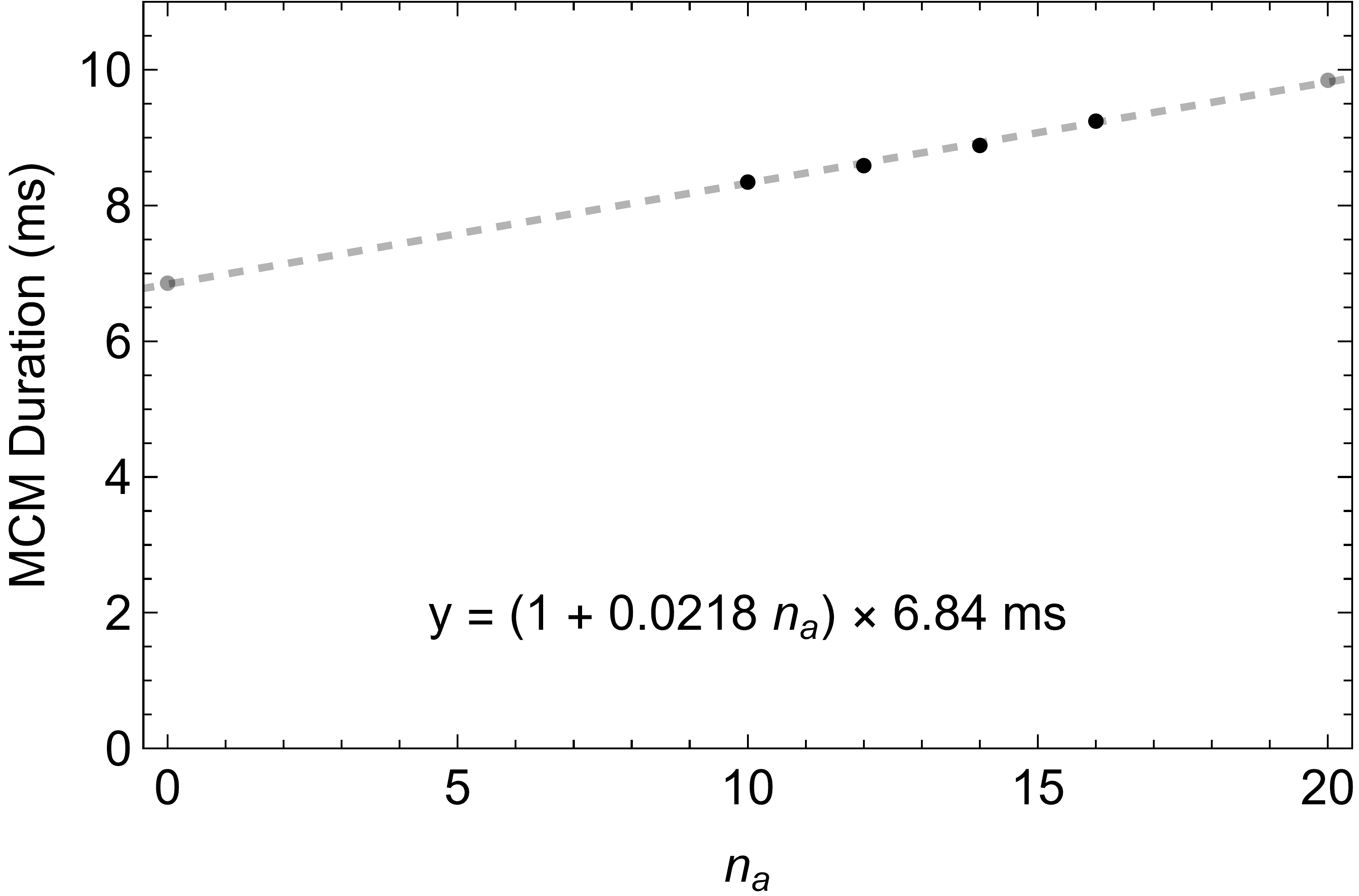}
\par\end{centering}
\caption{Duration of the MCM protocol as a function of $n_a$, the number of ancillae read out. The black points represent values of $n_a$ used for the codes presented in this work (see \cref{tab:NumAncillae}).}
\label{fig:MCMDurations}
\end{figure}

We now examine how the duration of the MCM protocol scales with the number of ancillae being read out ($n_a$). A major advantage of our OMG-based MCM technique is that an arbitrary set of ancillae can be read out in a single MCM round, but that operation is not completely parallel. The 1762 nm shelving/deshelving pulses and all operations within $\Sqb$ are performed in parallel with global beams, but the Raman gates used to transfer ancillae from either $\ket{0_D}$ or $\ket{1_D}$ to the $\Dqb$ intermediate state and back are applied sequentially by steerable 532 nm beams. In \cref{fig:MCMDurations}, we show the duration of the MCM protocol for values of $n_a$ from 0 to 20, corresponding to reading out up to half the chain. The linear fit demonstrates that this duration is dominated by the fixed, parallel operations, and reading out each additional ancilla only increases the total duration by approximately 2\%. Because the MCM duration scales so weakly with $n_a$, the MCM protocol can be practically considered to be a parallel operation.

\section{Comparing Physical and Logical Qubits}
\label{app:Physical Logical Qubit Comparison}
Here we define the noise channels $\Noiset$ for our physical and logical qubits, and we use them to compute explicit expressions for the qubit lifetimes that we report.
We will take the {\em process fidelity} $F_{\Noise}(t)$ (w.r.t. the identity) to mean the fidelity of a state after having evolved through the channel $\Noise$ for a period of time $t$, for input states averaged over the Haar measure~\cite{pedersen_fidelity_2007}.

We describe an idling atomic qubit as undergoing $\Tone$ and $\Ttwo$ relaxations~\cite{bloch_nuclear_1946}.
These relaxations are defined (over some time $t$) as the channel with Kraus set:
\begin{align}
K\in & \Bigg\{\left[\begin{array}{cc}
1 & 0\\
0 & e^{-t/\Ttwo}
\end{array}\right],\left[\begin{array}{cc}
0 & \sqrt{1-e^{-t/T_{1}}}\\
0 & 0
\end{array}\right],\nonumber\\
 & \left[\begin{array}{cc}
0 & 0\\
0 & \sqrt{1-e^{-2t/\Ttwo}e^{t/T_{1}}}e^{-t/2T_{1}}
\end{array}\right]\Bigg\}
\label{eq:T1T2Kraus}
\end{align}
This yields process fidelity and decay speed:
\begin{align}
  \FidAvg_{\text{phys}}(t)=&\frac{1}{2}+\frac{1}{6}(2e^{-t/\Ttwo} + e^{-t/\Tone})\nonumber \\
  D_{\text{phys}}(0)=&\frac{1}{3\Ttwo} + \frac{1}{6\Tone}\\
  \label{eq:FidPhys}
\end{align}
For our $\BaIon$ qubits, $\Tone$ is immeasurably long, so we take $1/\Tone=0$, which implies a physical qubit lifetime of $L_{\text{phys}} = 3\Ttwo$.

For our logical qubits, we define a general Pauli channel with Kraus set:
\begin{align}
  K & \in\left\{ \sqrt{\gamma_{0}}I\right\} \cup\left\{ \sqrt{\frac{1-e^{-\gamma_{\sigma}t}}{\left|P\right| + 1}}\sigma\,\bigg|\,\sigma\in P\right\} ,\\
  \text{where } & \gamma_{0}=\frac{1}{\left|P\right|+1}\left(1+\sum_{\sigma\in P}e^{-\gamma_{\sigma}t}\right)
  \label{eq:GeneralPauliChannel}
\end{align}
Here, $P\subseteq\{X, Y, Z\}$ is a set of single-qubit Pauli operators.
The corresponding process fidelity and decay speed (with $|P|=3$) are:
\begin{align}
  \FidAvg_{\log}(t)=&\frac{1}{3}+\frac{1}{6}\left(1+\sum_{\sigma\in P}e^{-\gamma_{\sigma}t}\right)\nonumber\\
  D_{\log}(0)      =&\frac{1}{6}\left(\sum_{\sigma\in P}\gamma_{\sigma}\right)
  \label{eq:FidLog}
\end{align}

To determine parameters $\gamma_{\sigma}$ of the general Pauli channel, we let the decay speed of $X$ and $Z$ eigenstates under this channel be equal to that observed in experimental data, as $t\to0$ hence:
\begin{align}
  \frac{1}{2 T_X} & =\frac{1}{4}\left(\gamma_{Y}+\gamma_{Z}\right)\\
  \frac{1}{2 T_Z} & =\frac{1}{4}\left(\gamma_{X}+\gamma_{Y}\right)
\end{align}
We remark that, because our memory experiments were performed only on $X$ and $Z$ eigenstates, parameters $\gamma_{\sigma}$ cannot all be uniquely determined from experimental data alone.
In~\cite{acharya_quantum_2025}, the channel is determined by also setting $\gamma_Y=0$, which maximizes $p_{L}^{(Y)}$ and gives a pessimistic estimate of lifetime.
Here we instead choose $\gamma_X = \gamma_Y$, which reflects the noise bias on our device (see \cref{tab:PauliNoise}), and agrees with what we observe in simulations of our syndrome circuits.
This yields the logical qubit lifetime:
\begin{align}
  \frac{1}{L_{\log}}=\frac{1}{3}\left(\frac{1}{2T_{Z}}+\frac{1}{T_{X}}\right)
\end{align}

\section{Physical $\Ttwo$ Measurement}
\label{app:Physical T2 Measurement}

\begin{figure}[h]
\begin{centering}
\includegraphics[width=7.5cm]{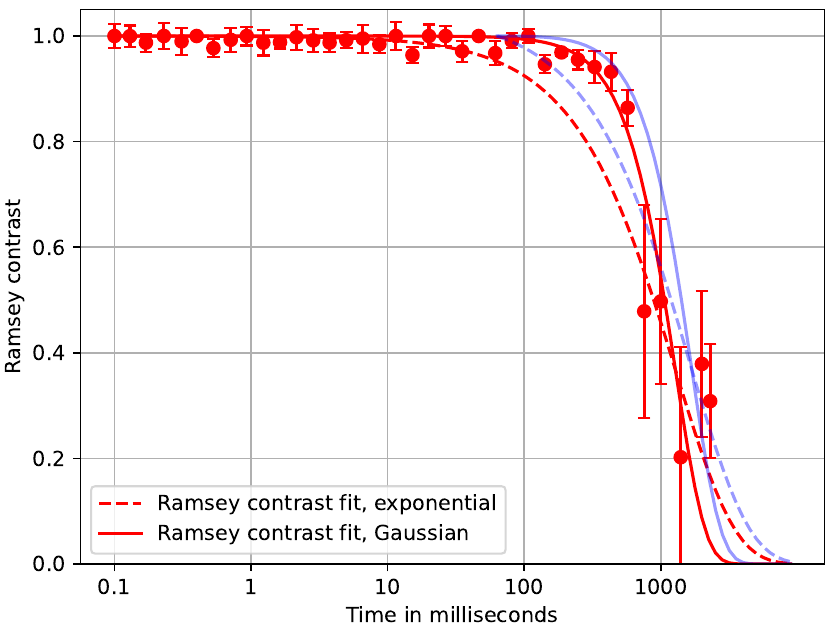}
\par\end{centering}
\caption{Ramsey measurement of $\Ttwo$ of physical qubits.
Fit curves are shown for both exponential $e^{-t/\Ttwo}$ (dashed) and Gaussian $e^{-(t/\Ttwo)^2}$ (solid) decays.
Also shown are similar curves but with the time constant set to $\Txsec$ for the GB$[[26,2,5]]$, as guides to the eye.}
\label{fig:T2star}
\end{figure}

The $\Ttwo$ coherence time of our system was characterized with a standard Ramsey experiment~\cite{ramsey_molecular_1950}.
\Cref{fig:T2star} shows the Ramsey contrast for our $\BaIon$ $\Sqb$ qubit, as a function of time.

We remark that the $\Ttwo$ phase decay we observe in our physical qubits are overwhelmingly driven by noise in the magnetic field environment, rather than by any $T_2$ decay that is fundamental to the hyperfine splitting itself.
Magnetic field noise acts over time on an initial preparation of $\ket{+}$, with a $Z$ error as follows:
\begin{align}
  \ket{+}\to\int_{0}^{\infty}\text{d}\omega f(\omega)e^{iZ\omega t}\ket{+}
  \label{eq:PhaseError}
\end{align}
Here, $f(\omega)$ is a spectral density function that describes the noise.

For common well-behaved $f(\omega)$, at times $t$ greater than the auto-correlation time implied by $f$, the phase decay behaves in accordance with the channel of \cref{eq:T1T2Kraus}.
On the other hand, at times when $t \Delta\omega\to 0$ (where $\Delta\omega$ is the width of distribution $f$), then the error described in \cref{eq:PhaseError} approximates a coherent or quasi-static evolution, and the survival of $\ket{+}$ is more closely described by $\cos^2(\alpha t)$ (for some parameter $\alpha$), than by the more familiar $e^{-t/\Ttwo}$.

\Cref{fig:T2star} shows data from a Ramsey experiment on our system.
Solid and dashed red lines are fits to data with $e^{-(t/\Ttwo)^2}$ and $e^{-t/\Ttwo}$ respectively, from which we obtain $\Ttwo=1.28\pm 0.16$~s.
Also shown are curves of similar forms in blue, but setting $\Ttwo=\Txsec$ for one logical qubit of the GB$[[26,2,5]]$ code, as guides to the eye.

Magnetic field noise on our system are observed to be very narrowly concentrated around $\omega\approx 1$~Hz and, to a lesser extent, $\omega\approx 60$~Hz, with correlation times greater than $1$s.
This results in the Gaussian decay form being a substantially better fit to our Ramsey data.
Furthermore we remark that the quality of the fits are generally poor especially at the decay tail ($t>1$s), because of partial revivals in contrast (again due to the quasi-static nature of our field noise) that prevent a clean monotonic decay to zero.
While this complicates precise estimates of our physical qubits' decay rate, it also suggests echoing ({\emph i.e.} dynamical decoupling) can dramatically improve lifetimes and logical performance.

\label{app:T2star}

\end{document}